\documentclass[a4paper,11pt]{amsart}
\usepackage{latexsym}
\usepackage{amsmath}
\usepackage{amssymb}
\usepackage{amsfonts}
\usepackage{mathrsfs}
\usepackage{mathabx}
\usepackage{tensor}
\usepackage{float}
\usepackage{esint}
\usepackage[utf8]{inputenc}
\usepackage[T1]{fontenc}   
\usepackage{graphicx}
\usepackage{color}
\allowdisplaybreaks 
\definecolor{Blue}{rgb}{0.,0.,1.}
\definecolor{Red}{rgb}{1.,0.,0.}
\definecolor{Green}{rgb}{0.,1.,0.}
\let\origmaketitle\maketitle
\def\maketitle{
  \begingroup
  \def\uppercasenonmath##1{} 
  \let\MakeUppercase\relax 
	\origmaketitle
  \endgroup
	}

\newcounter{smallarabics}
\newenvironment{arabicenumerate}
{\begin{list}{{\normalfont\textrm{(\arabic{smallarabics})}}}
  {\usecounter{smallarabics}\setlength{\itemindent}{0cm}
   \setlength{\leftmargin}{5ex}\setlength{\labelwidth}{4ex}
   \setlength{\topsep}{0.75\parsep}\setlength{\partopsep}{0ex}
   \setlength{\itemsep}{0ex}}}
{\end{list}}

\newcounter{smallroman}

\newcommand{\ben}{\begin{arabicenumerate}}  
\newcommand{\een}{\end{arabicenumerate}}

\def\init{\setcounter{equation}{0}}

\newtheorem{theoreme}{Theorem }[section]
\newtheorem{proposition}[theoreme]{Proposition}

\newtheorem{lemma}[theoreme]{Lemma}
\newtheorem{definition}[theoreme]{Definition}

\newtheorem{remark}[theoreme]{Remark}
\newtheorem{example}[theoreme]{Example}
\newcommand{\beq}{\begin{equation}}
\newcommand{\eeq}{\end{equation}}

\newcommand{\bex}{\begin{example}}
\newcommand{\eex}{\end{example}}
\def\bel{\begin{lemma}}
\def\eel{\end{lemma}}
\def\bet{\begin{theoreme}}
\def\eet{\end{theoreme}}
\def\bed{\begin{definition}}
\def\eed{\end{definition}}
\def\ber{\begin{remark}}
\def\eer{\end{remark}}

\newenvironment{notations}
{\begin{list}{{\normalfont\textrm{-}}}
  {\setlength{\itemindent}{0cm}
   \setlength{\leftmargin}{2ex}\setlength{\labelwidth}{4ex}
   \setlength{\topsep}{0.75\parsep}\setlength{\partopsep}{1ex}
   \setlength{\itemsep}{1ex}}
}
{\end{list}}

\def\rr{{\mathbb R}}

\def\cc{{\mathbb C}}
\def\nn{{\mathbb N}}

\def\part{{\rm par}}

\let\Im\relax
\let\Re\relax
\DeclareMathOperator{\Im}{Im}
\DeclareMathOperator{\Re}{Re}
\def\slim{{\rm s-}\lim}

\def\bar{\overline}

\def\cinf{C^\infty}
\def\c0inf{C_0^\infty}

\def\proof{
\noindent{\bf Proof.}\ \ }

\def\ch{{\rm ch}}
\def\th{{\rm th}}

\def\cV{{\mathcal V}}

\def\cS{{\mathcal S}}

\def\cD{{\mathcal D}}

\def\cW{{\mathcal W}}

\def\cf{C^\infty}

\def\i{{\rm i}}

\def\qed{$\Box$\medskip}

\def \p{ \partial}

\def\12{\frac{1}{2}}
\def\14{\frac{1}{4}}

\def\e{{\rm e}}

\def\bbbone{{\mathchoice {\rm 1\mskip-4mu l} {\rm 1\mskip-4mu l}
{\rm 1\mskip-4.5mu l} {\rm 1\mskip-5mu l}}}
\def\one{\bbbone}
\def\cH{{\mathcal H}}

\def\coinf{C_{\rm c}^\infty}

\def \p{ \partial}

\def\12{\frac{1}{2}}

\def\e{{\rm e}}

\def\cH{{\mathcal H}}

\def\bep{\begin{proposition}}
\def\eep{\end{proposition}}

\newcommand{\mat}[4]{\left(\begin{array}{cc}#1 &#2  \\ #3 &#4 \end{array}\right)}

\newcommand{\col}[2]{\left(\begin{array}{c}#1 \\#2\end{array} \right)}

\def\CARal{{\rm C\hskip 0.25 em \hbox{\raise 1.72 ex 
\hbox{$\scriptscriptstyle\rm al$}\kern -0.57 em A}R}}

\def\t{{\scriptscriptstyle\#}}
\def\otimesal{\mathop{\hbox{\raise 1.5 ex
  \hbox{$\scriptscriptstyle\rm al$}
\kern -0.92 em \hbox{$\otimes$}}}}
\def\oplusal{\mathop{\hbox{\raise 1.5 ex
  \hbox{$\scriptscriptstyle\rm al$}
\kern -0.92 em \hbox{$\oplus$}}}}
\def\Gammal{\hbox{\raise 1.68 ex 
\hbox{$\scriptscriptstyle\rm al$}\kern -0.50 em $\Gamma$}}
\def\Bal{\hbox{\raise 1.68 ex 
\hbox{$\scriptscriptstyle\rm  al$}\kern -0.50 em $B$}}
\def\CARal{{\rm C\hskip 0.25 em \hbox{\raise 1.72 ex 
\hbox{$\scriptscriptstyle\rm al$}\kern -0.57 em A}R}}
\def\t{{\scriptscriptstyle\#}}
\newcommand{\traa}[1]{\mskip-6mu\upharpoonright_{#1}}

\def\cE{{\mathcal E}}

\def\WFA{{\rm WF}_{\rm a}}
\makeatletter
\newcommand*{\defeq}{\mathrel{\rlap{%
                     \raisebox{0.3ex}{$\m@th\cdot$}}%
                     \raisebox{-0.3ex}{$\m@th\cdot$}}%
                     =}
\makeatother
\makeatletter
\newcommand*{\eqdef}{=\mathrel{\rlap{%
                     \raisebox{0.3ex}{$\m@th\cdot$}}%
                     \raisebox{-0.3ex}{$\m@th\cdot$}}%
                     }
\makeatother

\newcommand{\bea}{\begin{aligned}}
\newcommand{\beal}{\begin{array}{l}}
\newcommand{\eeal}{\end{array}}
\newcommand{\eea}{\end{aligned}}
\newcommand{\bec}{\begin{cases}}
\newcommand{\eec}{\end{cases}}

\def\cN{\mathcal{N}}
\def\wf{{\rm WF}}

\def\rx{{\rm x}}

\newcommand{\coo}[1]{T^{*}#1\backslash \zero}

\def\vol{{\rm vol}}

\DeclareMathOperator{\Dom}{Dom}
\DeclareMathOperator{\supp}{supp}

\def\tu{u}

\def\tcH{\cH}
\def\tK{K}

\def\tC{C}
\def\tgam{\gamma}

\def\mo{\mathscr{O}}
\def\ri{{\rm r}}
\def\le{{\rm l}}
\def\calde{Calder\'{o}n }
\def\wavefront{wave front }

\def\zero{{\rm\textit{o}}}
\DeclareMathOperator{\rs}{rs}

\newcommand{\NewBox}{{\raisebox{0.15ex}{$\Box$}}}

\begin{document}
\title[Analytic Hadamard states and  Calder\'{o}n projectors]{{\large Analytic Hadamard states, Calder\'{o}n projectors } \\ \vspace{0.06cm}   {\large and Wick rotation near
analytic Cauchy surfaces} }
\author{}
\address{Universit\'e Paris-Sud XI, D\'epartement de Math\'ematiques, 91405 Orsay Cedex, France}
\email{christian.gerard@math.u-psud.fr}
\author[]{\normalsize Christian \textsc{G\'erard} \& Micha{\l} \textsc{Wrochna}}
\address{Universit\'e Grenoble Alpes, CNRS, Institut Fourier, F-38000 Grenoble, France}
\email{michal.wrochna@univ-grenoble-alpes.fr}
\keywords{Quantum Field Theory on curved spacetimes, Hadamard states, \calde projector, Wick rotation}
\subjclass[2010]{81T13, 81T20, 35S05, 35S35}

\begin{abstract}
We consider the Klein-Gordon equation on analytic spacetimes with an analytic Cauchy surface. In this setting, we prove the existence of pure analytic Hadamard states. The proof is based on considering an elliptic operator obtained by Wick rotating the Klein-Gordon operator in a neighborhood of a Cauchy hyper\-surface. The Cauchy data of Hadamard two-point functions are constructed as Calder\'{o}n projectors (suitably genera\-lized if the hyper\-surface is non-compact) for the elliptic operator. 
\end{abstract}

\maketitle

\section{Introduction \& summary}\label{sec0}\init


\subsection{Analytic Hadamard condition} In Quantum Field Theory on  curved spacetimes, the \emph{Hadamard condition} plays a key role as a mean to select physically relevant states and as an ingredient in the renormalization of a priori ill-defined products of quantum fields. In the setup of the linear Klein-Gordon equation, it amounts to a condition on the $\cf$ \wavefront set of certain bi-solutions: the field's \emph{two-point functions}.

Presently, many techniques to construct two-point functions satisfying the Hadamard condition are available. This includes abstract proofs of existence on arbitrary globally hyperbolic spacetimes \cite{FNW,GW1}, as well as more explicit constructions on classes of spacetimes with good behavior at spatial infinity \cite{Ju,JS,GW1,GOW} and various other additional assumptions \cite{Ol,BT}. Furthermore, other strategies have been developed for spacetimes with specific asymptotic structures, in which case it is possible to have distinguished candidates for Hadamard two-point functions \cite{Mo,DMP1,DMP2,BJ,sanders,GW3,VW}, or to use global arguments \cite{BF,VW} (cf. \cite{GHV,GW4} for the related problem of constructing Feynman generalized inverses).

The situation is however dramatically different if one requires an analogue of the Hadamard condition with the $\cf$ \wavefront set replaced by the \emph{analytic \wavefront set} $\WFA$ (see Def. \ref{def:wfa}), assuming that the spacetime is analytic. The \emph{analytic Hadamard condition} was introduced by Strohmaier, Verch and Wollenberg, who have shown that whenever satisfied, it has remarkable consequences for the quantum field theory, as it implies the \emph{Reeh-Schlieder property} \cite{SVW}. This means that any vector in the Hilbert space can
be approximated arbitrarily well by acting on the vacuum with operations performed in any prescribed open region (see Subsect. \ref{ss:reeh}). Unfortunately, only  few examples of two-point functions were shown to satisfy the analytic Hadamard condition, namely the \emph{ground} and \emph{KMS states} on analytic stationary spacetimes with an analytic  Killing vector field \cite{SVW}. Moreover, the methods developed for the $\cf$ case do not seem to be directly useful in that respect, outside of the stationary case, as they rely on ingredients specific to the $\cf$ setting, such as spacetime deformation or variants of H\"ormander's pseudo\-differential calculus.

In the present work we fill this gap by providing a construction of analytic Hadamard two-point functions without assuming any symmetries of the spacetime. In the language of algebraic QFT,  we construct \emph{analytic Hadamard} (quasi-free) \emph{states} (see Sect. \ref{sec1.1}--\ref{sec1.3} for the relevant definitions), which are more precisely \emph{pure} ones (see Prop. \ref{purity} for a criterion formulated in terms of two-point functions). In this terminology, the main result can be stated as follows.

Let us recall that a hypersurface in a spacetime is \emph{Cauchy} if it  is intersected by every inextensible, causal  (i.e. non-spacelike) curve exactly
once.  

\begin{theoreme}\label{thm:main} Let $(M,g)$ be an analytic spacetime with an analytic space-like Cauchy hypersurface. Suppose $P$ is a differential operator of the form $P=-\NewBox_g + V$, where $V:M\to\rr$ is real analytic. Then there exists a pure analytic Hadamard state for $P$.
\end{theoreme}

\subsection{Outline of proof} The main steps in the proof of Thm. \ref{thm:main} can be summarized as follows.

First, we show that the problem can be reduced to a situation where the spacetime is replaced by a neighborhood of a Cauchy surface $\Sigma$ and the Lorentzian metric is of the form $-dt^2+h_t(y) dy^2$. Moreover, we argue that $t\mapsto h_t$ can be assumed without loss of generality to be a real analytic family of metrics on $\Sigma$ such that $h_0$ is complete. In this setup, the Klein-Gordon operator reads
 \[
P= \p_{t}^{2}+ r(t, y)\p_{t}+ a(t, y, \p_{y}),
\]
where $a$ is an elliptic differential operator of order $2$.

Next, we perform the {\em Wick rotation} in $t$, i.e. by means of the substitution  $t\eqdef\i s$ we get an elliptic operator
\[
K= - \p_{s}^{2}- \i r(\i s, y)\p_{s}+ a(\i s, y, \p_{y})
\]
defined on a neighborhood of $\{0\}\times\Sigma$. After possibly choosing a smaller neighborhood $\Omega$, we associate to $K$ a closed operator $K_{\Omega}$ by imposing Dirichlet boundary conditions on the boundary of $\Omega$  and we prove that $K_\Omega$ is inversible.

The construction is then an adaptation of a recent idea from \cite{HHI}, which consists in considering the \emph{\calde projectors} of $K_\Omega$ on the hypersurface $\{s=0\}$. While the \calde projectors belong to the standard toolbox of elliptic problems on manifolds with boundary (see e.g. \cite{Gr,H3}), their use in QFT in \cite{HHI} and in the present work is new. In addition to what can be found in the literature, here we also need to cope with issues related to the fact that $K$ is in general not formally self-adjoint, and the hypersurface $\{s=0\}$ might not be compact.

The rough idea is that for  solutions of $K_{\Omega}u=0$ in $\Omega^\pm=\{\Omega\cap \pm s>0\}$, one can consider their Cauchy data
\[
\gamma^\pm u = \col{u\traa{\Sigma}}{-\p_{s}u\traa{\Sigma}},
\]
where the trace on $\Sigma$ is understood as a limit from $\Omega^\pm$. 
  
Supposing for the moment that $\Sigma$ is compact, the space of all such data can be characterized as the range of a projector $C_\Omega^\pm$. It is well-known that these projectors can be constructed using the formula
\beq\label{eq:dlkfj}
C_{\Omega}^{\pm}\defeq  \mp \gamma^{\pm}K_{\Omega}^{-1} \gamma^{*}S,
\eeq
where $S$ is a suitable $2\times2$ matrix of multiplication operators and
\[
\gamma^{*}f= \delta(s)\otimes f_{0}+ \delta'(s)\otimes f_{1}, \ \ f= \col{f_{0}}{f_{1}}\in \coinf(\Sigma)^{2}.
\]
Now, while $C_\Omega^\pm$ are constructed in the Wick-rotated elliptic setting, we use them as the Cauchy data of bi-solutions $\Lambda^\pm$ for the original hyperbolic problem. The key property which allows us to conclude the analytic Hadamard condition for $\Lambda^\pm$ is:
\beq\label{eq:bwelk}
\forall f\in\cE'(\Sigma)^{2}, \ \ \WFA (U_{\Sigma}C_{\Omega}^{\pm}f)\subset \{\pm\tau\geq 0\},
\eeq
where $\tau$ is the covariable dual to $t$ and $U_{\Sigma} C_{\Omega}^{\pm}f$ is the unique solution of $P u=0$ with Cauchy data on $\Sigma$ equal to $C_{\Omega}^{\pm}f$. The general strategy in the proof of \eqref{eq:bwelk} is based on ideas due to Pierre Schapira and is to a large degree a special case of the analysis in \cite{Sch}. In our setting, the main step, (forgetting the space variables $x\in \Sigma$),  consists in constructing the analytic continuation of the function $(-K_{\Omega}^{-1} \gamma^{*}Sf)(s)$, and proving that the so-obtained holomorphic function has a boundary value which is precisely the distribution $(U_{\Sigma}C_{\Omega}^{+}f)(t)$. Using propagation of singularities theorems we show that it is sufficient to do so locally, and we give such local construction basing on theorems on representations of distributions as sums of boundary values of holomorphic functions.

To show that the pair of operators $C_{\Omega}^\pm$ can be used to define a state, one needs to check that it satisfies the identity (well-known in the compact case)
\beq\label{eq:bwelk2}
C_{\Omega}^+ + C_{\Omega}^-=\one,
\eeq
and a positivity condition (see Thm. \ref{turlututu} for the precise statement). It turns out that the latter can be proved by an argument reminiscent of \emph{reflection positivity} in Euclidean QFT. 

Finally, still supposing that $\Sigma$ is compact, the purity statement in Thm. \ref{thm:main} is merely a direct corollary of the operators $C_\Omega^\pm$ being projections. 

The case of $\Sigma$  non-compact is technically more involved. We show that formula \eqref{eq:dlkfj} still makes sense without the need of making extra assumptions on the geometry at spatial infinity, and \eqref{eq:bwelk}-\eqref{eq:bwelk2}  remain valid. However, purity is more subtle because it is not a priori clear if there is a suitable space on which $C_\Omega^\pm$ are projections. Using an approximation argument we show instead that $C_\Omega^\pm$ satisfy a weaker condition, which implies the purity statement nevertheless. The approximation argument is also used in the proof of the positivity properties of $C_\Omega^\pm$ in the non-compact case.

\subsection{Discussion and outlook} Once the Wick rotation is performed on the geometric level and the (sufficiently small) neighborhood $\Omega$ of $\{0\}\times \Sigma$ is chosen, our construction provides a canonical choice of analytic Hadamard two-point functions. 
Thus, in situations where a Cauchy surface is chosen and where the size of $\Omega$ is immaterial or under control, the construction assigns unambiguously a pair of Hadamard two-point functions to the spacetime. While that assignment is not expected to be locally covariant in the sense of \cite{BFV}, it could be of practical interest in the study of semi-classical Einstein equations nevertheless.  

An interesting issue is the relation of the present work to the construction of Hadamard states by pseudo-differential techniques in \cite{Ju,JS,GW1,GOW}. We expect that an alternative proof of \eqref{eq:bwelk} could be given using a `Wick-rotated' version of the parametrix in \cite{GOW}. Here, we chose not to use such arguments in order to avoid any assumptions on the geometry of the Cauchy surface at infinity. It is however worth stressing that the present work was triggered by the observation\footnote{This observation was kindly communicated to us by Francis Nier, to whom we are very grateful.} that the parametrix construction from \cite{GOW} is reminiscent  of parametrices for \calde projectors in elliptic problems. 

The construction in the present paper is also expected to preserve some symmetries that are not directly under control in techniques developed for the general $\cf$ case \cite{FNW,GW1}. This could be of particular merit e.g. in gauge theories.

It is worth pointing out that if the spacetime has special symmetries, an analogous construction with different boundary conditions can be particularly useful. This is for instance the case on spacetimes with a bifurcate Killing horizon, where a \calde projector corresponding to periodic boundary conditions can be used to construct the Hartle-Hawking-Israel state \cite{HHI}.

An interesting perspective would be the extension of the present construction to setups where the spacetime has a boundary. This would require a good understanding of \calde projectors on manifolds with corners.  

\subsection{Plan of the paper} The main part of the paper is structured as follows.

In Sect. \ref{sec1} we review standard definitions and results on Hadamard states and analytic Hadamard states. In Sect. \ref{sec2} we perform the Wick rotation and study various properties of the elliptic operator $K_\Omega$. The associated (generalized) \calde projectors $C^\pm_\Omega$ are defined and analyzed in Sect. \ref{seccald}. We prove in particular that they can be used to define \emph{pure} quasi-free states. We also briefly discuss the special case of an ultra-static metric. Sect. \ref{sec5} is devoted to the proof of the analytic Hadamard condition; it includes an introduction to the analytic \wavefront set and its basic properties. An auxiliary lemma is deferred to the appendix.

\subsection{Notation}\label{notgen} Throughout the paper we adopt the following notations and conventions.

\begin{notations}
\item We write $A\Subset B$ if $A$ is relatively compact in $B$.
\item If $X,Y$ are sets and $f:X\to Y$ we write  $f: X \xrightarrow{\sim}Y$ if $f$ is
bijective.  If $X, Y$ are equipped with topologies, {we write $f:X\to Y$ if the map is continuous, and $f: X \xrightarrow{\sim}Y$ if it is a homeomorphism.}
\item  The domain of a closed, densely defined operator  $a$  will be denoted by $\Dom a$. 
\end{notations}

\section{Quantum Klein-Gordon fields}\init\label{sec1}
\subsection{Klein-Gordon fields}\label{sec1.1}
In this section we review classical results about quasi-free states for free Klein-Gordon quantum fields   on a globally hyperbolic spacetime,  see e.g. \cite{BGP, DG,KM,HW} for textbook accounts and recent reviews. 
We will use the complex formalism, based on charged fields. 
 \subsubsection{Bosonic quasi-free states}\label{sec1.1.1}
Let $\cV$ be a complex vector space, $\cV^{*}$ its anti-dual and let us denote by $L_{\rm h}(\cV, \cV^{*})$ the space of hermitian sesquilinear forms on $\cV$. We denote by $\bar{v}_{1}\cdot q v_{2}$ the evaluation of $q\in L(\cV, \cV^{*})$ on  $v_{1}, v_{2}\in \cV$.

  A pair $(\cV,q)$ consisting of a complex vector space $\cV$ and a non-degenerate hermitian form $q$ on $\cV$ will be called a {\em phase space}.  Denoting by $\cV_{\rr}$ the real form of $\cV$, i.e. $\cV$ considered as a real vector space, $(\cV_{\rr}, \Im q)$ is a real symplectic space.
  
The 
{\em   CCR $*$-algebra} ${\rm CCR}(\cV,q)$  is the abstract $*-$algebra generated by $\one$ and  elements $\psi(v), \psi^{*}(v)$ for $v\in \cV$, subject to relations:
  \[
  \begin{array}{l}
  \psi(v+ \lambda w)= \psi(v)+ \overline{\lambda}\psi(w), \\[2mm]
    \psi^{*}(v+ \lambda w)= \psi^{*}(v)+\lambda\psi^{*}(w), \\[2mm]
  [\psi(v), \psi(w)]= [\psi^{*}(v), \psi^{*}(w)]=0,  \ \ [\psi(v), \psi^{*}(w)]=  \bar{v} q w \one,\\[2mm]
  \psi(v)^{*}= \psi^{*}(v), \ \ \lambda\in \cc, v, w\in \cV.
  \end{array}
 \]
Equivalently, ${\rm CCR}(\cV, q)$ is generated by $\one$ and  $\phi(v)$ for $v\in \cV$, where the (abstract) {\em real fields} $\phi(v)$ are  defined by
 \[
\phi(v)\defeq \frac{1}{\sqrt{2}}(\psi(v)+ \psi^{*}(v)),
\]
and
\[
[\phi(v_{1}),\phi(v_{2})]= \i v_{1}\cdot \Im qv_{2}, \ v_{1}, v_{2}\in \cV.
\]
 A (gauge invariant) quasi-free state $\omega$ on ${\rm CCR}(\cV,q)$ is  entirely characterized by  its 
\emph{complex covariances}  $\Lambda^\pm\in L_{\rm h}(\cV,\cV^*)$  defined by
\beq\label{eq:lambda}
\bar{v}\cdot\Lambda^+ w \defeq \omega\big(\psi(v)\psi^*(w)\big), \quad \bar{v}\cdot\Lambda^- w \defeq \omega\big(\psi^*(w)\psi(v)\big), \quad v,w\in \cV,
\eeq
since gauge invariance implies that
\[
 \omega\big(\psi(v)\psi(w)\big)=  \omega\big(\psi^{*}(v)\psi^*(w)\big)=0, \ v, w\in \cV.
\]
Note that  $\Lambda^{\pm}\geq 0$ and   $\Lambda^+ - \Lambda^- = q$ by the canonical commutation relations.
Conversely if   $\Lambda^\pm$  are Hermitian forms on $\cV$ such that  
\begin{equation}
\label{positon}
\Lambda^+ - \Lambda^- = q, \ \ \Lambda^\pm\geq 0,
\end{equation}
 then there is a unique   (gauge invariant) quasi-free state $\omega$ such that (\ref{eq:lambda}) holds, see e.g. \cite[Sect. 17.1]{DG}. One can associate to $\omega$ the pair of operators $c^{\pm}\in L(\cV)$:
\beq\label{defodefo}
c^{\pm}\defeq  \pm q^{-1}\circ \Lambda^{\pm}.
\eeq
The properties \eqref{positon} become then:
\beq\label{defidefi}
c^{+}+ c^{-}=\one, \ \ c^{\pm*}q= q c^{\pm}, \ \ \pm q c^{\pm}\geq 0.
\eeq
In the real formalism one has:
\[
\omega(\phi(v_{1})\phi(v_{2}))=  v_{1}\cdot \eta v_{2}+ \frac{\i}{2}v_{1}\Im qv_{2},
\]
where  the symmetric form $\eta\in L_{\rm s}(\cV, \cV^{\t})$ is called the {\em real covariance} of $\omega$.

We recall a well-known characterization of {\em pure states} (see e.g. \cite{KW} in the real case  and \cite[Prop. 7.1]{GOW} for another equivalent characterization).
\begin{proposition}\label{purity}
 The state $\omega$ with covariances $\Lambda^{\pm}$ is {\em pure} iff:
 \beq\label{arbo}
\bar{v}_{1}\cdot (\Lambda^{+}+ \Lambda^{-})v_{1}= \sup_{v_{2}\in \cV, v_{2}\neq 0}\frac{|\bar{v}_{1}\cdot q v_{2}|^{2}}{\bar{v}_{2}\cdot (\Lambda^{+}+ \Lambda^{-})v_{2}}, \ \ \forall v_{1}\in \cV.
\eeq
\end{proposition}
\proof Consider $\cV$ as a real vector space, equipped with the symplectic form $\sigma= \Im q$. Then from \cite[Subsect. 2.3]{GW1} we know that the real covariance of $\omega$ is $\eta= \12 \Re(\Lambda^{+}+ \Lambda^{-})$. From \cite{KW} we know that $\omega$ is a pure state iff
\[
v_{1}\cdot \eta v_{1}= \frac{1}{4}\sup_{v_{2}\neq 0}\frac{|v_{1}\cdot\Im q v_{2}|^{2}}{v_{2}\cdot \eta v_{2}}.
\]
Using that $\eta=  \12 \Re(\Lambda^{+}+ \Lambda^{-})$ and that $q$ is sesquilinear, this is equivalent to \eqref{arbo}. \qed

\subsubsection{Klein-Gordon fields}\label{sec1.1.2} We adopt the convention
that a \emph{spacetime} is a Hausdorff, paracompact, connected, time orientable
smooth Lorentzian ma\-ni\-fold equipped with a time orientation.

Let $(M, g)$ be a globally hyperbolic spacetime, i.e. a spacetime possessing a smooth spacelike Cauchy hypersurface, and let
\[
P= - \nabla^{a}\nabla_{a}+ V(x), \ \ V\in \cinf(M, \rr)
\]
be a Klein-Gordon operator on $(M, g)$. 
 
 We denote by $G_{\rm ret/adv}$ the retarded/advanced inverses for $P$ (see e.g. \cite{BGP}) and by $G\defeq  G_{\rm ret}- G_{\rm adv}$  the Pauli-Jordan commutator.  We set
\beq\label{pscal.e1}
(u|v)_{M}\defeq  \int_{M}\bar{u}v \,d\vol_{g},  \ \ u, v\in\coinf(M),
\eeq
and we will use $(\cdot| \cdot)_{M}$ to identify sesquilinear forms on $\coinf(M)$ with linear operators from $\coinf(M)$ to $\cinf(M)$.
We set:
\beq\label{defo}
\cV\defeq \frac{\coinf(M)}{P\coinf(M)}, \quad \overline{[u]} \cdot Q[ v]\defeq \i (u| G v)_{M}.
\eeq
It is well-known that $(\cV, Q)$ is a phase space. 

Let now $\Sigma$ be a smooth spacelike Cauchy hypersurface  for $(M, g)$ and $\cV_{\Sigma}= \coinf(\Sigma;\cc^{2})$. We equip $\cV_{\Sigma}$ with the scalar product
\beq\label{pscal.e3}
(f|g)_{\Sigma}\defeq \int_{\Sigma}\overline{f}_{0}g_{0}+ \overline{f}_{1}g_{1}d\sigma_{\Sigma},
\eeq
and again, we use $(\cdot| \cdot)_{\Sigma}$ to identify sesquilinear forms on $\coinf(\Sigma; \cc^{2})$ with linear operators from $\coinf(\Sigma; \cc^{2})$ to $\cinf(\Sigma; \cc^{2})$.
Let us set
\[
\rho_{\Sigma}u\defeq  \begin{pmatrix}u\traa{\Sigma} \\ \i^{-1}\partial_n u\traa{\Sigma}\end{pmatrix},
\]
where $n$ is the future unit normal to $\Sigma$  and:
\begin{equation}
\label{pscal.e4}
q_{\Sigma}=\mat{0}{1}{1}{0}
\end{equation}
 It is   well known that  the map:
\[
\rho_{\Sigma}G: (\cV, Q)\to (\cV_{\Sigma}, q_{\Sigma})
\]
is pseudo-unitary, i.e.
\[
\overline{\rho_{\Sigma} G u} \cdot q_{\Sigma}\rho_\Sigma Gv =  \overline{[u]} \cdot Q[ v], \ \ u,v\in\coinf(M).
\]
One can use equivalently
either of the two above phase spaces.    The ${\rm CCR}*-$algebra associated to either $(\cV, Q)$ or $(\cV_{\Sigma}, q_{\Sigma})$ will be simply denoted by ${\rm CCR}(P)$.
\subsubsection{Spacetime two-point functions}\label{sec1.1.3}

We will use for the moment the phase space $(\cV, Q)$ defined in (\ref{defo}). Let us introduce the conditions:
\begin{equation}
\label{eq:titu}
\begin{array}{rl}
i)&\Lambda^{\pm}: \coinf(M)\to\cinf(M),\\[2mm]
ii)&\Lambda^{\pm}\geq 0 \hbox{ for }(\cdot| \cdot)_{M} \hbox{ on }\coinf(M),\\[2mm]
iii)& \Lambda^{+}-\Lambda^{-}= \i G,\\[2mm]
iv)& P\Lambda^{\pm} = \Lambda^{\pm}P=0,\\[2mm]
v)& (u| \Lambda^{\pm}u)_{M}\geq 0, \ \ \forall u\in \coinf(M).
\end{array}
\end{equation}
As explained above, we set with a slight abuse of notation:
\[
\overline{[u]}\cdot\Lambda^{\pm}[v]\defeq  (u|\Lambda^{\pm}v)_{M}, \ \ [u], [v]\in \frac{\coinf(M)}{P\coinf(M)}.
\]
If  (\ref{eq:titu}) hold,   then $\Lambda^{\pm}$  define a pair of complex covariances on the phase space $(\cV, Q)$ defined in (\ref{defo}), hence define a unique quasi-free state  on ${\rm CCR}(P)$.
\begin{definition}
 A pair  of maps  $\Lambda^{\pm}: \coinf(M)\to\cinf(M)$  satisfying (\ref{eq:titu})   will be called a pair of {\em  space\-time two-point functions}. 
\end{definition}
\subsubsection{Cauchy surface two-point functions}\label{sec1.1.4}
We will need a version of two-point functions acting on Cauchy data for  $P$ instead of  test functions on $M$. 

Let us introduce the assumptions:
\begin{equation}
\label{eq:titucauchy}
\begin{array}{rl}
i)&\lambda^{\pm}_{\Sigma}: \coinf(\Sigma; \cc^{2})\to\cinf(\Sigma; \cc^{2}), \\[2mm]
ii)&\lambda_{\Sigma}^{\pm}\geq 0 \hbox{ for }(\cdot| \cdot)_{\Sigma},\\[2mm]
iii)& \lambda_{\Sigma}^{+}- \lambda_{\Sigma}^{-}= q_{\Sigma}.
\end{array}
\end{equation}
\begin{definition}\label{defcs2pt}
 A pair of maps $\lambda_{\Sigma}^{\pm}$ satisfying (\ref{eq:titucauchy}) will be called a pair of {\em Cauchy surface two-point functions}.
\end{definition}

The following proposition is shown in \cite{GW2} in a more general situation.
\begin{proposition}\label{minusu}
The maps:
 \[
 \bea\lambda_{\Sigma}^{\pm}\mapsto \Lambda^{\pm}\defeq  (\rho_{\Sigma} G)^{*}\lambda_{\Sigma}^{\pm}(\rho_{\Sigma} G),\\[2mm]
 \Lambda^{\pm}\mapsto\lambda_{\Sigma}^{\pm}\defeq (\rho_{\Sigma}^{*}q_{\Sigma})^{*} \Lambda^{\pm} (\rho_{\Sigma}^{*}q_{\Sigma})
 \eea
\]
are bijective and inverse from one another. Furthermore, $\lambda_{\Sigma}^{\pm}$ are Cauchy surface two-point functions iff $\Lambda^\pm$ are spacetime two-point functions.
\end{proposition}

\subsection{ The Hadamard condition}\label{sec1.2}
We now recall the celebrated {\em Hadamard condition} in its microlocal formulation.

We start by recalling some standard notation.

\begin{notations}
\item For $x\in M$ we denote by $V_{x\pm}\subset T_{x}M$ the future/past solid lightcones and by $V_{x\pm}^{*}\subset T_{x}^{*}M$ the dual cones $V_{x\pm}^{*}=\{\xi\in T^{*}_{x}M :\  \xi\cdot v>0, \ \forall v\in V_{x\pm}, \ v\neq 0\}$. We write
\[
\xi\rhd 0\ (\hbox{resp. }\xi\lhd 0)\,\hbox{ if }\,\xi\in V_{x+}^{*} \ (\hbox{resp. } V_{x-}^{*}).
\]
\item We denote by $\zero$ the zero section of $T^*M$. 
\item For $X= (x, \xi)\in T^{*}M\backslash\zero$ we denote by $p(X)= \xi\cdot g^{-1}(x)\xi$ the principal symbol of $P$ and by
$\cN= p^{-1}(0)\cap \coo{M}$ the characteristic manifold of $P$. If $H_{p}$ is the Hamiltonian vector field of $p$,
integral curves of $H_{p}$ in $\cN$ are called {\em bicharacteristics}. $\cN$ splits into the {\em upper/lower energy shells}
\[
\cN= \cN^{+}\cup \cN^{-}, \ \ \cN^{\pm}=\cN\cap \{\pm\xi\rhd 0\}.
\]
\item If $\Gamma\subset T^{*}M\times T^{*}M$ we set
\[
\Gamma'=\left\{\big((x_{1}, \xi_{1}), (x_{2}, \xi_{2})\big): \big((x_{1}, \xi_{1}), (x_{2}, -\xi_{2})\big)\in \Gamma\right\}.
\]
\item If  $u\in\cD'(M)$, the \emph{\wavefront set} of $u$ is denoted by $\wf(u)$ and is a closed conic subset of $T^{*}M\backslash\zero$.
\end{notations}

By the Schwartz kernel theorem, we can  identify $\Lambda^\pm$ with a pair of distributions $\Lambda^\pm(x, x')\in \cD'(M\times M)$. One is especially interested in the subclass of \emph{Hadamard states}, which are subject to a condition on the \textit{\wavefront set} of $\Lambda^\pm(x, x')$.  
\begin{definition}\label{def:hadamard} A quasi-free state $\omega$ on ${\rm CCR}(P)$ is a \emph{Hadamard state} if its covariances $\Lambda^\pm$ satisfy $\Lambda^\pm: C_{\rm c}^\infty(M)\to\cf(M)$ and
\beq\label{hadamard}
\wf(\Lambda^\pm)'\subset \cN^\pm\times\cN^\pm.
\eeq
\end{definition}
{  This form of the Hadamard condition is taken from \cite{SV,hollands}. The original formulation in terms of \wavefront sets is due to Radzikowski \cite{R}, who showed its equivalence with older definitions \cite{KW}.

A fundamental result is the following existence theorem of Hadamard states, which was proved by Fulling, Narcovich and Wald in \cite{FNW}:
\begin{theoreme}[{\cite{FNW}}]\label{fnw}
 Let $P$ be a Klein-Gordon operator on a globally hyperbolic spacetime $(M,g)$. Then there exist pure, quasi-free Hadamard states for $P$.
\end{theoreme}
 The proof of this result proceeds by constructing an interpolating metric $\tilde{g}$ and a Klein-Gordon operator $\tilde{P}$ which equal $g$, $P$ in the far future of some Cauchy surface $\Sigma$ and equal  $g_{\rm us}$, $P_{\rm us}$ in the far past of $\Sigma$, where $g_{\rm us}$ is some ultrastatic metric and $P_{\rm us}$ is the associated Klein-Gordon operator with a constant mass. Transporting the vacuum state for $g_{\rm us}$ to the far future of $\Sigma$ by the evolution of $\tilde{P}$ yields a  state for $P$. The Hadamard condition on $g$ is then concluded from the Hadamard property of the vacuum state for  $g_{\rm us}$  using the propagation of singularities theorem and the time-slice property of Klein-Gordon fields.
 
\subsection{The analytic Hadamard condition}\label{sec1.3}
In \cite{SVW}, Strohmaier, Verch and Wollenberg introduced the notion of {\em analytic Hadamard states}, obtained from Def. \ref{def:hadamard} by replacing the $C^{\infty}$ \wavefront set   $\wf$ by the  {\em analytic \wavefront set} $\WFA$. 

The definition of  the analytic \wavefront set $\WFA(u)$ of  a distribution $u\in \cD'(N)$,  for $N$  a real analytic manifold, will be recalled in Subsect. \ref{sec5.2}.

The basic results of microlocal analysis, such as microlocal ellipticity or propagation of singularities theorems, require to consider differential operators with analytic coefficients when one wants to study for example the analytic \wavefront set of solutions.

Therefore the notion of analytic Hadamard states is restricted to {\em analytic spacetimes} $(M, g)$, i.e.  real analytic manifolds $M$ equipped with a real analytic Lorentzian metric $g$. The Klein-Gordon operator $P$ is now
\begin{equation}
\label{e1.anal}
P= - \nabla^{a}\nabla_{a}+ V(x), \hbox{ where }V: M\to \rr \hbox{ is real analytic}. 
\end{equation}
We will call a Klein-Gordon operator as above an {\em analytic Klein-Gordon operator}.
\begin{definition}
  A quasi-free state on ${\rm CCR}(P)$ is an \em{analytic Hadamard state} if its spacetime covariances $\Lambda^\pm$ satisfy 
\beq\label{hadamardanal}
\WFA(\Lambda^\pm)'\subset \cN^\pm\times\cN^\pm.
\eeq
\end{definition}

 In \cite{SVW} the analytic Hadamard condition is also defined for more general states on ${\rm CCR}(P)$ by extending the microlocal spectrum condition of Bru\-netti, Freden\-hagen and K\"{o}hler \cite{BFK} on the $n-$point functions  to the analytic case. 
 
 It is shown in \cite[Prop. 6.2]{SVW} that  for quasi-free states, the analytic microlocal spectrum condition is equivalent to the  following real version of \eqref{hadamardanal}.
 
Let $\phi(u)= \frac{1}{\sqrt{2}}(\psi(u)+ \psi^{*}(u))$   for $u\in \coinf(M)$ {\em real},  and let $\omega_{2}\in \cD'(M\times M)$ be the (real) two-point function of $\omega$ defined by
\[
\omega(\phi(u)\phi(v))= \langle \omega_{2}, u\otimes v\rangle,\ \ u, v\in \coinf(M; \rr).
\]
Then the $C^{\infty}$ (resp. analytic) Hadamard condition is equivalent to:
\begin{equation}
\label{real-anal}
\wf(\omega_{2})'\ (\hbox{resp. }\WFA(\omega_{2})')\subset\cN^{+}\times \cN^{+}.
\end{equation}
  In \cite[Remark 3.3]{GW1} it is shown that  for the  $C^{\infty}$ \wavefront set  \eqref{real-anal} is equivalent to \eqref{hadamard}. The same argument is valid for the analytic \wavefront set.
  
  It has been shown by Radzikowski \cite{R} that the covariances of any two Hadamard states coincide modulo a smooth kernel. The same is true for analytic Hadamard states:

\begin{proposition}
 Let $\omega, \tilde{\omega}$ be two analytic Hadamard states for some analytic Klein-Gordon operator $P$. Then 
 $\Lambda^{\pm}- \tilde{\Lambda}^{\pm}$ have analytic kernels.
\end{proposition}
\proof  Let $R^{\pm}= \Lambda^{\pm}- \tilde{\Lambda}^{\pm}$. Since $\Lambda^{+}- \Lambda^{-}=  \tilde{\Lambda}^{+}-  \tilde{\Lambda}^{-}=\i G$, we have $R^{+}=- R^{-}$. On the other hand from \eqref{hadamardanal} we have $\WFA(R^{\pm})'\subset \cN^{\pm}\times \cN^{\pm}$ hence $\WFA(R^{+})'\cap \WFA(R^{-})'=\emptyset$. Since $R^{-}= - R^{+}$ this implies that $\WFA(R^{\pm})'= \emptyset$ hence $R^{\pm}$ have analytic kernels. \qed
\subsection{The Reeh-Schlieder property}\label{ss:reeh}
In \cite{SVW}, Strohmaier, Verch and Wollenberg proved an important consequence of the analytic Hadamard condition.
\begin{theoreme}[\cite{SVW}]\label{theosvw}
An analytic Hadamard state on ${\rm CCR}(P)$ satisfies the {\em Reeh-Schlieder} property.
\end{theoreme}
 We first recall how the  Reeh-Schlieder property is defined. If $\omega$ is a state on ${\rm CCR}(P)$ we denote by $(\cH_{\omega}, \pi_{\omega}, \Omega_{\omega})$ the GNS triple associated to $\omega$.
\begin{definition}
 A state $\omega$ on ${\rm CCR}(P)$ satisfies the {\em Reeh-Schlieder property} if for any open set $U\subset M$ the space 
 \[
 {\rm Vect}\big\{\pi_{\omega}\left(\textstyle\prod_{1}^{p}\psi^{*}(u_{i})\textstyle\prod_{1}^{q}\psi(v_{j})\right)\Omega_{\omega}: \ p, q\in \nn, \ u_{i}, v_{j}\in \coinf(U)\big\}
 \] is dense in $\cH_{\omega}$.
\end{definition}

The proof of Thm. \ref{theosvw} in \cite{SVW} relies on two ingredients: the first is the use of \wavefront set for Hilbert space valued distributions.  The second is the fact that the 
 analytic \wavefront set  $\WFA (u)$ of  $u\in \cD'(M)$ has  deep relations with the support $\supp u$. An example of such a relation is the {\em Kashiwara-Kawai theorem}, which we state below for illustration, and which plays a key role in \cite{SVW}.

If $F\subset M$ is a closed set, the {\em  normal set} $N(F)\subset T^{*}M\backslash \zero$ is the set of $(x^{0}, \xi^{0})$ such that $x^{0}\in F$, $\xi^{0}\neq 0$, and there exists  a real function $f\in C^{2}(M)$ such that $df(x^{0})= \xi^{0}$ or $df(x^{0})= -\xi^{0}$ and $F\subset \{x: f(x)\leq f(x^{0})\}$. Note that $N(F)\subset T^{*}_{\p F}M$.

The {\em Kashiwara-Kawai theorem} (see e.g. \cite[Thm. 8.5.6']{H2}) states that
\begin{equation}
\label{ekawai}
N(\supp u)\subset \WFA (u)\ \ \forall u\in \cD'(M).
\end{equation}

\subsection{Existence of analytic Hadamard states on analytic spacetimes}
It is not an easy task to construct analytic Hadamard states. The main problem is that the Fulling-Narcowich-Wald deformation argument \cite{FNW} used to prove Thm. \ref{fnw} does not apply anymore as the interpolating metric $\tilde{g}$ is not real analytic  (though  at least a weaker form of the Reeh-Schlieder property can be obtained, and an abstract existence argument can be given for states satisfying the full Reeh-Schlieder property, see \cite{sandersanalytic}). 

The only examples of analytic Hadamard states known so far are the {\em vacua (ground states)} and {\em KMS states} on analytic, stationary spacetimes with an analytic  Killing vector field, see \cite[Thm. 6.3]{SVW}.

Our main result, Thm. \ref{thm:main}, provides a general existence proof for any analytic Klein-Gordon operator $P$ on an analytic spacetime with an analytic, spacelike Cauchy hypersurface. 



Before introducing the main new ingredients of our construction, let us recall two standard facts which are useful to construct Hadamard states, both in the $C^{\infty}$ and analytic case, the first one relying on propagation of singularities and the second on conformal transformations.

\begin{proposition}\label{prop-reduc1}
 Let $(M, g)$ be a globally hyperbolic analytic spacetime, $P$ an analytic Klein-Gordon operator on $(M,g)$, $\omega$ a quasi-free state  for $P$ and $\Sigma$ a Cauchy hypersurface for $(M, g)$.
 
If  its covariances $\Lambda^{\pm}$ satisfy  the analytic Hadamard condition \eqref{hadamardanal} over some neighborhood $U$ of $\Sigma$ then they satisfy \eqref{hadamardanal} everywhere.
  \end{proposition}
  \proof 
Let   $\phi_{s}$, $s\in \rr$ be  the Hamiltonian flow of $p$.
 By \eqref{eq:titu} {\it iv)}, microlocal ellipticity   and   propagation of singularities  in the analytic case (see \cite[Thm. 3.3']{kawai} or \cite[Thm. 7.1]{H4}), we know that  if $(X_{1}, X_{2})\in \WFA(\Lambda^{\pm})'$, then $X_{1}, X_{2}\in \cN$ and $(\phi_{s_{1}}(X_{1}), \phi_{s_{2}}(X_{2}))\in \WFA(\Lambda^{\pm})'$ for all $s_{1}, s_{2} \in \rr$.
  Since $\Sigma$ is a Cauchy hypersurface, there exists $s_{1}, s_{2}$ such that $\phi_{s_{i}}(X_{i})\in T^{*}U$, hence $\phi_{s_{i}}(X_{i})\in \cN^{\pm}$, hence $X_{i}\in \cN^{\pm}$.
\qed

\begin{proposition}\label{prop-reduc2}
 Let $(M, g)$ be a globally hyperbolic analytic spacetime with a Cauchy hypersurface $\Sigma$.
 
  Suppose that there exists a neighborhood $U$ of $\Sigma$ in $M$ and an analytic function $c: U\to ]0, +\infty[$ such that any analytic Klein-Gordon operator on $(U, c^{2}g)$ has a pure  analytic Hadamard state. Then any analytic Klein-Gordon operator on $(M, g)$ has a pure  analytic Hadamard state.
\end{proposition}
  \proof We write $P$ as $-\Box_{g}+ \frac{n-2}{4(n-1)}R_{g}+ W$ with $W$ real analytic. By conformal invariance of $-\Box_{g}+ \frac{n-2}{4(n-1)}R_{g}$, setting $\tilde{g}= c^{2}g$ we have:
  \[
  \tilde{P}\defeq c^{-n/2-1}P c^{n/2-1}= - \Box_{\tilde g} +\frac{n-2}{4(n-1)}R_{\tilde g}+ \tilde{W},
  \]
  where $\tilde{W}= c^{-2}W$.  If $\tilde{G}$ is the Pauli-Jordan commutator for $\tilde{P}$ we have $G= c^{n/2-1}\tilde{G}c^{-n/2-1}$. It follows that if $\tilde{\Lambda}^{\pm}$ are the covariances of some  quasi-free 
  state $\tilde{\omega}$ for $\tilde{P}$, then $\Lambda^{\pm}=c^{n/2-1}\tilde{\Lambda}^{\pm}c^{-n/2-1}$ are the covariances of  some   quasi-free state $\omega$ for $P$.  
  
If $\tilde{\omega}$ is a pure state then so is $\omega$.  Indeed, denoting by $(\tilde{\cV}, \tilde{Q})$    the classical phase space for $\tilde{P}$, the map 
  \[
 T:\tilde{\cV}\ni [\tilde{u}]\mapsto  [c^{n/2+1}\tilde{u}]\in\cV
  \]
  is pseudo-unitary from $(\tilde{\cV}, \tilde{Q})$ to $(\cV, Q)$, and $\omega$ is simply the pushforward of $\tilde{\omega}$ by $T$.
  
 If $\tilde{\omega}$ is an analytic Hadamard state for $\tilde{P}$, $\tilde{\Lambda}^{\pm}$ satisfy \eqref{hadamardanal} over $U$, and so do $\Lambda^{\pm}$ since $c$ is analytic. By  Prop. \ref{prop-reduc1} $\Lambda^{\pm}$ satisfy \eqref{hadamardanal} over $M$, hence $\omega$ is an analytic Hadamard state. \qed

 \section{ Wick rotation on analytic spacetimes}\label{sec2}\init
 In this section we perform the Wick rotation in Gaussian normal coordinates. If $\Sigma$ is a spacelike Cauchy hypersurface in $(M, g)$, the Klein-Gordon operator $P$  is written as
 \[
P= \p_{t}^{2}+ r(t, y)\p_{t}+ a(t, y, \p_{y}),
\]
if $(t, y)$ are Gaussian normal coordinates respective to $\Sigma$. By analyticity one can perform the {\em Wick rotation} $t\eqdef\i s$ near $s= 0$ and consider the {\em Wick rotated operator}
\[
K= - \p_{s}^{2}- \i r(\i s, y)\p_{s}+ a(\i s, y, \p_{y}),
\]
which is an elliptic differential operator defined on some neighborhood $V$ of $\{0\}\times \Sigma$ in $\rr\times \Sigma$. To define a proper  {\em inverse} $K^{-1}$, we need some realization of $K$ as an unbounded operator.  As turns out, the natural way is to realize $K$ by imposing  {\em Dirichlet boundary conditions} on a sufficiently small neighborhood $\Omega$ of $\{0\}\times \Sigma$ in $\rr\times \Sigma$.
\subsection{Gaussian normal coordinates}
Let us consider a globally hyperbolic analytic spacetime $(M, g)$ with an analytic spacelike Cauchy hypersurface $\Sigma$.

Using  Gaussian normal coordinates to $\Sigma$, we  obtain neighborhoods $U$ of $\{0\}\times \Sigma$ in $\rr\times \Sigma$ and $U'$ of $\Sigma$ in $M$ and an isometric diffeomorphism
\beq\label{b5.0}
\chi: (U, -dt^{2}+ h_{t}(y)dy^{2})\to (U', g),
\eeq
where  if  $K\Subset \Sigma$  and  $\epsilon>0$  is such that $]-\epsilon, \epsilon[\times K\subset U$, then $]-\epsilon, \epsilon[\ni t\mapsto h_{t}(y)dy^{2}$ is a $t-$dependent Riemannian metric on $K$. In particular  $h_{0}(y)dy^{2}$ is the Riemannian metric induced by $g$ on $\Sigma$.

By the Cauchy-Kowalevski theorem, it follows from the fact that $(M, g)$ and $\Sigma$ are analytic, that $\chi: U\to U'$ is analytic, and that $ U\ni (t, y)\mapsto h_{t}(y)dy^{2}$ is an analytic $(2, 0)$ tensor.  

It will turn out convenient later on to assume that
\begin{equation}
\label{b5.1}
\hbox{the Riemannian manifold }(\Sigma, h_{0})\hbox{ is complete}.
\end{equation}
Let us explain how to reduce ourselves to this situation.

It is  known, see e.g. \cite{Kan}, that there exists a real analytic function  $c: \Sigma\to ]0, +\infty[$ such that $c^{2}h_{0}$ is complete on $\Sigma$. 

We extend $c$ to $U\subset \rr\times \Sigma$ by $c(t, y)= c(y)$, push it to $U'$ by $\chi$ and consider the Lorentzian metric $\tilde{g}= c^{2}g$. Clearly $(U', \tilde{g})$ is globally hyperbolic with $\Sigma$ as a spacelike Cauchy hypersurface.

By Prop. \ref{prop-reduc2}, to construct  analytic Hadamard states for some analytic Klein-Gordon operator $P$ on $(M, g)$, it suffices to perform the construction on $(U, \tilde{g})$, for some conformally rescaled analytic Klein-Gordon operator $\tilde{P}$. Denoting  $U$ by $M$ and  $\tilde{g}$  again by $g$, we can hence assume without loss of generality that \eqref{b5.1} holds.

\subsection{Klein-Gordon operator in Gaussian normal coordinates}
 Denoting   the operator $\chi^{*}P$ by $P$ again, we obtain that
 \begin{equation}
 \label{b5.2}
 P= \p_{t}^{2}+ r(t, y)\p_{t}+ a(t, y, \p_{y}),
 \end{equation}
 where
 \[
 r(t, y)= |h_{t}(y)|^{-\12}\p_{t}|h_{t}(y)|^{\12},
 \]
 and
 \[
 a(t, y, \p_{y})= |h_{t}(y)|^{-\12}\p_{j}|h_{t}(y)|^{\12}h_{t}^{jk}(y)\p_{k}+ V(t, y).
 \]
 The operator $a(t, y, \p_{y})$ is selfadjoint for the scalar product
 \[
 (u|v)= \int_{U}\bar{u}v|h_{t}|^{\12}dydt.
 \]
 \subsubsection{Reduction of $P$}\label{redoto}
 It is possible to reduce oneself to the case where $r(t, y)\equiv 0$. In fact if
 \beq\label{e3.1}
d(t, y)= |h_{t}(y)|^{1/4}|h_{0}(y)|^{-1/4},
\eeq
 we see that 
\[
d:L^{2}(U, |h_{t}|^{\12}dtdy)\ni u \mapsto d u\in L^{2}(U, |h_{0}|^{\12}dtdy)
\]
 is unitary and 
\begin{equation}
\label{e3.2}
P_{0}\defeq dPd^{-1}= \p_{t}^{2}+a_{0}(t, y, \p_{y}),
\end{equation}
for
\beq\label{e3.2connerie}
a_{0}(t, y, \p_{y})= d(t, y)a(t, y, \p_{y})d^{-1}(t, y)- \frac{r^{2}(t, y)}{4} - \12 \p_{t}r(t,y),
\eeq
which is selfadjoint for the scalar product
\beq\label{pscalo}
(u|v)_{0}= \int_{U}\bar{u}v|h_{0}|^{\12}dydt.
\eeq
\subsection{Wick rotated operator}\label{sec3.1}
 The function $t\mapsto r(t, \cdot )$ and the differential operator $t\mapsto a (t, y, \p_{y})$ extend holomorphically in $t$ in  a neighborhood $W$ of $\{0\}\times \Sigma$ in $\cc\times \Sigma$.  We can moreover assume that $W$ is small enough such that for each $\alpha\in \rr$ the functions $U\ni (t, y)\mapsto |h_{t}(y)|^{\alpha}$ extend holomorphically  in $t$ to $W$. In particular $d(t, y)$ defined in \eqref{e3.1} extends holomorphically  in $t$ to $W$. 
 
 We  define the {\em Wick rotated operator}:
 \beq\label{b5.4}
K= -\p_{s}^{2}- \i r(\i s , y)\p_{s}+ a(\i s, y, \p_{y}), \ (s, y)\in V,
 \eeq
where $V$  is a neighborhood of $\{0\}\times \Sigma$ in $\rr\times \Sigma$.  By possibly replacing it by a smaller neighborhood, we can assume that $V$ is invariant under the reflection $(s,y)\mapsto (-s, y)$.

The operator $K$ is obtained from $P$ by the substitution $t\to \i s$. It has analytic coefficients in $(s, y)$ on $V$.
 
 \subsubsection{Reduction of $K$}\label{sec2.2.1}
 The reduction of $P$ in  \ref{redoto} can be similarly carried out for $K$. In fact let us set:
 \begin{equation}
 \label{b5.5}
 \hat{h}_{s}(y)= (h_{\i s}^{*}h_{\i s})^{\12}(y), \ \ (s, y)\in U.
 \end{equation}
  which is positive definite.  Note that from $h_{t}(y)= h_{t}(y)^{*}$, we obtain $h_{\i s}(y)^{*}= h_{- \i s}(y)$ hence:
  \begin{equation}
  \label{ekito}
 |\hat{h}_{s}(y)|= |\hat{h}_{-s}(y)|, \  \ (s, y)\in U.
  \end{equation}
 We also set:
  \begin{equation}
\label{e5.5}
\hat{d}(s, y)= |h_{\i s}(y)|^{1/4}|h_{0}(y)|^{-1/4}= d(\i s, y)
\end{equation}
  for $d$ defined in \eqref{e3.1}.  We see that
    \beq\label{e4.00}
\hat{d}:L^{2}(U, |\hat{h}_{s}|^{\12}dyds)\ni u \mapsto d u\in L^{2}(U, |h_{0}|^{\12}dyds)
\eeq
 is unitary and from  \eqref{e3.2}  we obtain that:
 \begin{equation}
 \label{e3.5}
 \hat{d}K\hat{d}^{-1}\eqdef K_{0}= - \p_{s}^{2}+ a_{0}(\i s, y, \p_{y}).
 \end{equation}

From the selfadjointness of $a_{0}(t, y, \p_{y})$ we obtain that
\[
a_{0}(\i s, y, \p_{y})^{*}= a_{0}(- \i s, y, \p_{y})
\]
on $L^{2}(U, |h_{0}|^{\12}dyds)$.
\subsection{Some preparations}
The operators $K$ or $K_{0}$ are elliptic differential operators,  but have for the moment no realizations as unbounded operators with some concrete domain. We will fix such a realization by introducing {\em Dirichlet boundary conditions} on the boundary of some open set $\Omega\subset V$. In this subsection we collect some properties obtained from a convenient choice of $\Omega$.

\begin{lemma}\label{lem5.1}
 Let  $\Sigma= \bigcup_{i\in \nn}U_{i}$ be a covering of $\Sigma$ with $U_{i}\Subset \Sigma$. Then there exist $T_{i}$, $i\in \nn$ such that  for  all $(s, y)\in ]-T_{i}, T_{i}[\times U_{i}$ and all $v, v_{1}, v_{2}\in \cc T_{y}\Sigma$ one has:
 \beq\label{b5.6}
 \begin{array}{rl}
 i)&\12 \bar{v} \cdot h_{0}(y)v\leq \bar{v}\cdot \Re \cdot h_{\i s}(y)v\leq \frac{3}{2}\bar{v}  \cdot h_{0}(y)v,\\[2mm]
 ii)& |\bar{v} \cdot \Im h_{\i s}(y)v|\leq \12\bar{v}  \cdot h_{0}(y)v, \\[2mm]
 iii)& |\bar{v}_{1} \cdot h_{\i s}(y)v_{2}|\leq 2 (\bar{v}_{1}  \cdot h_{0}(y)v_{1})^{\12}(\bar{v}_{2} \cdot h_{0}(y)v_{2})^{\12}\\[2mm]
 iv)& \12 \leq |\hat{d}(s, y)|\leq1, \\[2mm]
 v)& \nabla \hat{d}(s, y)\cdot h_{0}(y)\nabla \hat{d}(s, y)\leq 1. \end{array}
  \eeq
\end{lemma}
 \proof  {\it i)} and {\it ii)} are obvious and imply {\it iii)}.  {\it iv)} and {\it v)} follow from $\hat{d}(0, y)\equiv 1$. \qed

Let $\Omega\subset U$  be an open neighborhood of $\{0\}\times \Sigma$ with a smooth boundary. For $s\in \rr$ we denote by $H^{s}_{\rm c}(\Omega)$, $H^{s}_{\rm loc}(\Omega)$ the compactly supported and local Sobolev spaces of order $s$. We denote by $H^{1}_{0}(\Omega)$ the closure of $\coinf(\Omega)$ for the norm
\[
\|u\|^{2}_{H^{1}(\Omega)}= \int_{\Omega}(|\p_{s}u|^{2}+ \p_{j}\bar{u}h^{jk}_{0}\p_{k}u+ |u|^{2})|h_{0}|^{\12}dyds.
\]

Note that if $(U_{i}, T_{i})_{i\in \nn}$ are as in Lemma \ref{lem5.1} and $\Omega\subset \bigcup_{i\in \nn} ]-T_{i}, T_{i}[\times U_{i}$, it follows from Lemma \ref{lem5.1} {\it iv)} and {\it v)} that
\begin{equation}
\label{b5.7}
\hat{d}: H^{1}_{0}(\Omega)\xrightarrow{\sim} H^{1}_{0}(\Omega).
\end{equation}

In Prop. \ref{p4.1} below the space $L^{2}(\Omega, |\hat{h}_{s}(y)|^{\12}dsdy)$ is denoted by $L^{2}(\Omega)$.

If $A$ is a closed operator, its resolvent set is denoted by $\rs(A)$.

\begin{proposition}\label{p4.1}
 Let $Q_{\Omega}$  be the sesquilinear form:
  \[
Q_{\Omega}(v, u)= (v| Ku)_{L^{2}(\Omega)}, \ \ \Dom Q_{\Omega}= \coinf(\Omega).
\]
Then there exists an $\Omega$ as above such that:
  \ben
  \item $\Omega$ is invariant under $i:(s, y)\mapsto (-s, y)$;
 \item $Q_{\Omega}$ and $Q_{\Omega}^{*}$ are closeable on $L^{2}(\Omega)$;
\item   their closures $\bar{Q_{\Omega}}$, $\bar{Q_\Omega^{*}}$  are  sectorial, with domain   $H^{1}_{0}(\Omega)$;
\item the closed operators $K_{\Omega}$, $K^{*}_{\Omega}$ associated to $\bar{Q_{\Omega}}$, $\bar{Q_\Omega^{*}}$  satisfy  $0\in \rs(K_{\Omega})$, $0\in \rs(K^{*}_{\Omega})$;
\item $K^{*}_{\Omega}$ is the adjoint of $K_{\Omega}$.
 \een
 \end{proposition}
\proof 
 Let us  denote $L^{2}(V, |h_{0}|^{\12}(y)dsdy)$  by $L^{2}_{0}(V)$ and similarly $L^{2}(\Sigma, | h_{0}|^{\12}dy)$ by $L_{0}^{2}(\Sigma)$.  We  first consider the  sesquilinear form 
 \[
Q_{0}(v, u)= (\p_{s}v|\p_{s}u)_{L_{0}^{2}(V)}+ (v|a_{0}(\i s)u)_{L_{0}^{2}(V)}, \Dom Q_{0}= \coinf(\Omega).
 \]
 Let  $\Sigma= \bigcup_{i\in \nn}U_{i}$ a covering of $\Sigma$ and $1= \sum_{i\in \nn}\chi_{i}^{2}(y)$, $\chi_{i}\in\coinf(U_{i})$ a  subordinate partition of unity on $\Sigma$.  Let us set 
 \[
 b(s)= \Re a_{0}(\i s), \ \ c(s)= \Im a_{0}(\i s), \ \ k_{0}= -\Delta_{h_{0}}+1,
 \]
 where the real and imaginary parts are computed w.r.t. the scalar product in $L^{2}_{0}(\Sigma)$.
  We have 
 \beq\label{e4.1}
 \begin{array}{l}
 b(s)= \sum_{i\in \nn}\chi_{i}b(s)\chi_{i}+m(s), \ \ m(s)= \12 \sum_{i}[\chi_{i}, [\chi_{i}, b(s)]],\\[2mm]
 c(s)= \sum_{i\in \nn}\chi_{i}c(s)\chi_{i}+n(s), \ \ n(s)= \12 \sum_{i}[\chi_{i}, [\chi_{i}, c(s)]],\\[2mm]
 k_{0}= \sum_{i\in \nn}\chi_{i}k_{0}\chi_{i}+m_{0}, \ \ m_{0}= \12 \sum_{i}[\chi_{i}, [\chi_{i}, k_{0}]].
\end{array}
 \eeq
Moreover,  $b(s)$ is a second order elliptic operator, formally selfadjoint on $L_{0}^{2}(\Sigma)$, with principal symbol $\eta\cdot \Re h_{\i s}^{-1}(y)\eta$.  By Lemma \ref{lem5.1} {\it i)} and {\it ii)}  there exist $c_{i}, T_{i}>0$ such that:
 \begin{equation}
\label{e4.0}
\begin{array}{rl}
\frac{1}{2}\chi_{i}k_{0}\chi_{i}- \chi_{i}c_{i}\chi_{i}\leq\chi_{i} b(s)\chi_{i}\leq \frac{3}{2}\chi_{i}k_{0}\chi_{i}+ \chi_{i}c_{i}\chi_{i}, \\[2mm]
-\frac{1}{2}\chi_{i}k_{0}\chi_{i}- \chi_{i}c_{i}\chi_{i}\leq\chi_{i} c(s)\chi_{i}\leq \frac{1}{2}\chi_{i}k_{0}\chi_{i}+ \chi_{i}c_{i}\chi_{i}
\end{array}
\end{equation}
 on $L_{0}^{2}(\Sigma)$ for $|s|\leq T_{i}$.  Since $m(s), n(s), m_{0}$ are  multiplication operators on $\Sigma$, there exist constants $c_{i}'$ such that
 \[
 \chi_{i}(| m|(s)+|n|(s)+ |m_{0}|)\chi_{i}\leq c_{i}'\chi_{i}^{2}, \hbox{ on }L^{2}_{0}(\Sigma),  \hbox{ for }|s|\leq T_{i}'.
 \]
It follows that 
 \beq\label{e4.2}
| m|+|n|+ |m_{0}|\leq \sum_{i\in \nn}c_{i}'\chi_{i}^{2}, 
\eeq 
on $L_{0}^{2}(\bigcup_{i\in \nn}U_{i}\times ]-S_{i}, S_{i}[),$ if $0<S_{i}<T_{i}'$.
 By \eqref{e4.1}, \eqref{e4.0}, \eqref{e4.2} we obtain:
 \beq\label{e4.2b2}
\begin{array}{l}
\12 k_{0}- \sum_{i}(c_{i}+ \12c_{i}')\chi_{i}^{2}\leq b\leq \frac{3}{2} k_{0}+ \sum_{i}(c_{i}+ \frac{3}{2}c_{i}')\chi_{i}^{2},\\[2mm]
-\12
k_{0} - \sum_{i}(c_{i}+ \12c_{i}')\chi_{i}^{2}\leq c\leq \12 k_{0}+ \sum (c_{i}+ c_{i}')\chi_{i}^{2},
\end{array}
\eeq
on $L_{0}^{2}(\bigcup_{i\in \nn}U_{i}\times ]-S_{i}, S_{i}[),$ if $0<S_{i}<\min(T_{i},T_{i}')$.

By the Poincaré inequality we can find $T''_{i}>0$ such that
 \beq\label{e4.3}
 (\p_{s}u| \p_{s}u)_{L^{2}_{0}(V)}\geq( 2c_{i}+ 2c'_{i}+1)(u| u)_{L^{2}_{0}(V)}, \ \ u\in \coinf(]-T''_{i}, T''_{i}[\times\Sigma).
 \eeq
Let now $S_{i}= \min (T_{i}, T'_{i}, T''_{i})$ and $\Omega\subset\bigcup_{i\in \nn}]-S_{i}, S_{i}[\times U_{i}$ an open  neighborhood of $\{0\}\times \Sigma$ with a smooth boundary and $i(\Omega)=\Omega$. We have by \eqref{e4.3}:
\beq\label{e4.3bb}
\bea
(\p_{s}u| \p_{s}u)_{L^{2}_{0}(V)}&= \sum_{i\in \nn}(\p_{s}\chi_{i}u| \p_{s}\chi_{i}u)_{L^{2}_{0}(V)}\\[2mm]
&\geq ( 2c_{i}+ 2c_{i}'+1)(u| \chi_{i}^{2}u)_{L^{2}_{0}(V)}, \ \ u\in \coinf(\Omega),
\eea
\eeq
since $\supp u\subset \Omega$ implies $\supp \chi_{i}u\subset ]-T''_{i}, T''_{i}[\times \Sigma$. 
Let us denote by $Q_{\rm ref}$ the hermitian form
\[
Q_{\rm ref}(v, u)= (\p_{s}v| \p_{s}u)_{L^{2}(V)}+ (v| k_{0}u)_{L^{2}(V)}, \ \ \Dom Q_{\rm ref}= \coinf(\Omega).
\]
 we obtain from \eqref{e4.2b2}, \eqref{e4.3bb}:
\beq\label{e4.4}
\begin{array}{l}
\frac{1}{2}Q_{\rm ref}(u, u)\leq \Re Q_{0}(u, u)\leq \frac{3}{2}Q_{\rm ref}(u, u),\\[2mm]
|\Im Q_{0}(u, u)|\leq \12 \Re Q_{\rm ref}(u, u), \ \ u\in \coinf(\Omega).
\end{array}
\eeq
The form $Q_{\rm ref}$ is closeable, strictly positive and its closure $\overline{Q}_{\rm ref}$  has domain $H^{1}_{0}(\Omega)$.  From \eqref{e4.4} we obtain that $Q_{0}$ is closeable on $\coinf(\Omega)$, and its closure $\overline{Q}_{0}$ has domain $H^{1}_{0}(\Omega)$ and is sectorial. By \cite[VI.2.1]{K}  the associated operator $K_{\Omega,0}$ is  closed, sectorial with  $0\in \rs(K_{\Omega, 0})$. 

The same is true of the form $Q_{0}^{*}$ and if $K^{*}_{\Omega,0}$ is the operator associated to its closure, we have $K^{*}_{\Omega,0}= (K_{\Omega})^{*}$ by \cite[Thm. VI.2.5]{K}.

Next, we set $K_{\Omega}\defeq \hat{d}^{-1}K_{\Omega, 0}\hat{d}$ with domain $\hat{d}^{-1}\Dom K_{\Omega,0}$.   Using \eqref{e4.00} and \eqref{e3.5} we see that $Q_{\Omega}, Q_{\Omega}^{*}$ are closeable, with associated operators $K_{\Omega}, K^{*}_{\Omega}$ and $K^{*}_{\Omega}= \hat{d}^{-1}K^{*}_{\Omega, 0}\hat{d}= (K_{\Omega})^{*}$.  \qed

We end this section with a lemma which states that away from $\p\Omega$, $K_{\Omega}^{-1}$ is given by a pseudodifferential operator on $\Omega$ of order $-2$. 

We denote by $\Psi^{m}_{\rm c}(\Omega)$ the space of classical, properly supported  pseudodifferential operators of order $m\in \rr$, see e.g. \cite{Sh}, and by $\cW^{-\infty}(\Omega)$ the space of {\em smoothing operators}, i.e.  linear operators on $\Omega$ with  smooth distributional kernels.  We set  $\Psi^{m}(\Omega)= \Psi^{m}_{\rm c}(\Omega)+ \cW^{-\infty}(\Omega)$ and 
 \beq\label{eq:pdo}
 \Psi^{\infty}_{({\rm c})}(\Omega)= \bigcup_{m\in \rr}\Psi^{m}_{(\rm c)}(\Omega).
 \eeq

\begin{lemma}\label{l4.1}
 Let $\varphi_{1}, \varphi_{2}\in \coinf(\Omega)$ with $\varphi_{1}= 1$ near $\supp \varphi_{2}$. Then there exist $Q\in \Psi^{-2}_{\rm c}(\Omega)$  such that
 \[
\varphi_{1}K_{\Omega}^{-1}\varphi_{2}= \varphi_{1}Q\varphi_{2}+ R_{-\infty}, \ \ R_{-\infty}\in \cW^{-\infty}(\Omega).
\]
\end{lemma}
\proof  From Lemma \ref{lem5.1} we obtain that $|\eta\cdot h_{\i s}^{-1}(y)\eta|\geq c_{0}  \eta\cdot h_{0}^{-1}(y)\eta$ for $(s, y)\in \Omega$, $\eta\in T^{*}_{y}\Sigma$, hence $K$ is elliptic on $\Omega$. It is well known that $K$ admits a properly supported {\em parametrix}, i.e. some  $Q\in \Psi^{-2}_{{\rm c}}(\Omega)$ such that:
\[
K Q- \one, QK-\one\in \cW^{-\infty}(\Omega).
\] 
We can moreover assume that $(1- \varphi_{1})Q\varphi_{2}=0$ hence:
\[
K Q\varphi_{2}= \varphi_{2}+ \varphi_{1}R_{-\infty}\varphi_{2}, \ \ R_{-\infty}\in \cW^{-\infty}(\Omega). 
\]
Since $\varphi_{1}=0$ near $\p \Omega$, this implies that
\[
\varphi_{1}Q\varphi_{2}=Q\varphi_{2}= K_{\Omega}^{-1}(\varphi_{2}+ \varphi_{1}R_{-\infty}\varphi_{2})= \varphi_{1}K_{\Omega}^{-1}\varphi_{2}+  \varphi_{1}K_{\Omega}^{-1}\varphi_{1}R_{-\infty}\varphi_{2}.
 \]
 By elliptic regularity $K_{\Omega}^{-1}: H^{s}_{\rm c}(\Omega)\to H^{s+2}_{\rm loc}(\Omega)$,  hence $ \varphi_{1}K_{\Omega}^{-1}\varphi_{1}R_{-\infty}\varphi_{2}\in \cW^{-\infty}(\Omega)$.  This completes the proof. \qed

\section{\calde  projectors and Hadamard states}\init\label{seccald}
In this section we construct  the {\em \calde projectors} $C_{\Omega}^{\pm}$ associated to $K_{\Omega}$. We show that  they define a pure quasi-free state for  the analytic Klein-Gordon operator $P$.
\subsection{Notation}\label{notato}
Let us fix an open set $\Omega\subset V$ such that Lemma \ref{lem5.1} and Prop. \ref{p4.1} hold. 
 If $\Omega_{1}\subset \Omega$ is open and $F(\Omega)\subset \cD'(\Omega)$ is a space of distributions, we denote by $\bar{F}(\Omega_{1})\subset \cD'(\Omega_{1})$ the space of {\em restrictions} of elements of $F(\Omega)$ to $\Omega_{1}$. 

It is well known that any $u\in \overline{\cD'}(\Omega_{1})$  has a unique extension $e u\in \cD'(\Omega)$ with $eu=0$ on $\Omega\backslash \Omega_{1}^{\rm cl}$. We will apply this to  the open sets:
\[
\Omega^{\pm}\defeq  \Omega\cap \{\pm s>0\}.
\]
For example elements of $\overline{\cinf}(\Omega^{\pm})$ are  functions in $\cinf(\Omega^{\pm})$ which extend smoothly across $s=0$, while  elements of $\overline{\coinf}(\Omega^{\pm})$ have additionally a compact support in $\overline{\Omega^{+}}$   and   vanish near $\p \Omega^{\pm}\cap\{\pm s>0\}$ (but not necessarily near $\Sigma= \p\Omega^{\pm}\cap \{s= 0\}$). Similarly elements of $\overline{H^{1}_{0}}(\Omega^{\pm})$ are functions in $H^{1}(\Omega^{\pm})$ which vanish on $\p \Omega^{\pm}\cap\{\pm s>0\}$, but not necessarily on $\Sigma$.

 We denote by $r^{\pm}: \cD'(\Omega)\to \overline{\cD'}(\Omega^{\pm})$, resp.
   $e^{\pm}:  \overline{\cD'}(\Omega^{\pm})\to \cD'(\Omega)$ the operators of restriction, resp. extension by $0$.  

We define the trace operator by
 \[
\gamma u= \col{u\traa{\Sigma}}{-\p_{s}u\traa{\Sigma}}, \ \ u\in \cinf(\Omega).
  \]
We denote by $\gamma^{\pm}$ the analogous trace operators defined on $\overline{\cinf}(\Omega^{\pm})$.  

We set
\[
\cH^{(\pm)}= L^{2}(\Omega^{(\pm)}, |\hat{h}_{s}|^{\12}dsdy), \ \ \cH_{0}^{(\pm)}=  L^{2}(\Omega^{(\pm)}, |h_{0}|^{\12}dsdy), 
\]
and 
\[
\cS= L^{2}(\Sigma, |h_{0}|^{\12}dy)\otimes \cc^{2}.
\]
We denote by $\gamma^{*}: \cE'(\Sigma)^{2}\to \cD'(\Omega)$ the formal adjoint of $\gamma: \coinf(\Omega)\to \coinf(\Sigma)^{2}$ when  $\coinf(\Omega)$, resp. $\coinf(\Sigma)^{2}$ is equipped with the scalar products of $\cH$, resp. $\cS$. We have:
 \beq\label{e3.16}
 \gamma^{*}f= \delta(s)\otimes f_{0}+ \delta'(s)\otimes f_{1}, \ \ f= \col{f_{0}}{f_{1}}\in \coinf(\Sigma)^{2}.
 \eeq
 This follows from the fact that $\p_{s}|\hat{h}_{s}(y)|^{\12}$ vanishes at $s=0$, because of \eqref{ekito}.

 \subsection{Some identities}
Let us set:
\begin{equation}
\label{alofa}
\begin{array}{l}
R= \mat{\one}{0}{ \p_{s}\hat{d}(0, y)}{\one}, \ \ q=\mat{0}{\one}{\one}{0}, \\[3mm]
 S= \mat{2\i \p_{t}d(0, y)}{-\one}{\one}{0}, \ \ S_{0}= \mat{0}{-\one}{\one}{0}.
\end{array}
\end{equation}
The following identities are straightforward to check:
\begin{equation}
\label{e3.12}
\gamma^{(\pm)}\circ \hat{d}= R \circ \gamma^{(\pm)},  \ \ R^{*} qR= q, \ \ R^{*}S_{0}R=S.
 \end{equation}
 
The lemma  below is proved by  direct computations, using  the form of $K_{0}$ in \eqref{e3.5} and integration by parts.
\begin{lemma}\label{l3.1}
Let $u\in \overline{\cinf}(\Omega^{\pm}) , v \in \overline{\coinf}(\Omega^{\pm})$. Then:
\beq\label{alato}
 (v |K_{0}u )_{\cH_{0}^{\pm}}- (K_{0}^{*}v | u )_{\cH_{0}^{\pm}}= \pm (\gamma^{\pm}v |S_{0}\gamma^{\pm}u )_{\cS}.
 \eeq
 \beq\label{alatoto}
 \bea
 &(v| K_{0}u)_{\cH_{0}^{\pm}}+ (K_{0}v| u)_{\cH^{\pm}_{0}}\\
 &= 2(\p_{s}v| \p_{s}u)_{\cH^{\pm}_{0}}+ (v| (a_{0}(\i s)+ a_{0}(\i s)^{*})u)_{\cH_{0}^{\pm}}\mp (\gamma^{\pm}v| q\gamma^{\pm}u)_{\cS}.
 \eea
  \eeq
 \end{lemma}

The following proposition will be needed in the proof of Thm. \ref{turlututu} below. It  is superfluous if the Cauchy hypersurface $\Sigma$ is compact.

For $\chi\in\coinf(\Sigma)$ we set $\| \nabla \chi\|_{\infty}= \sup_{y\in \Sigma}|\p_{j}\chi(y)h_{0}^{jk}(y)\p_{k}\chi(y)|^{\12}$.

\begin{proposition}\label{arlito} Let  $\chi$ be a compactly supported Lipschitz function on $(\Sigma, h_{0})$. There exists $c_{0}>0$ such that for all $u, v\in \overline{H^{1}_{0}}(\Omega^{\pm})$:
 \[
 \begin{array}{rl}
  |(\chi v| [a_{0}(\i s), \chi]u)_{\cH_{0}^{\pm}}|
  \leq c_{0} (\| \chi\|_{\infty}\| \nabla\chi\|_{\infty}+ \| \nabla\chi\|^{2}_{\infty})\| v\|_{H^{1}(\Omega^{\pm})}\| u\|_{H^{1}(\Omega^{\pm})}, \\[2mm]
   |(\chi v| [a_{0}(\i s)^{*}, \chi]u)_{\cH_{0}^{\pm}}|
  \leq c_{0} (\| \chi\|_{\infty}\| \nabla\chi\|_{\infty}+ \| \nabla\chi\|^{2}_{\infty})\| v\|_{H^{1}(\Omega^{\pm})}\| u\|_{H^{1}(\Omega^{\pm})}.
 \end{array}
\]
 \end{proposition}
\proof Let us set $\hat{d}_{s}(y)= \hat{d}(s, y)$. It suffices to prove the result for  $a_{0}(\i s)$, using that $a_{0}(\i s)^{*}= a_{0}(- \i s)$.  Moreover by density we can assume that $u,v\in \overline{\coinf}(\Omega^{\pm})$. From \eqref{e3.5} we obtain that $[a_{0}(\i s), \chi]= \hat{d}_{s}[a(\i s), \chi]\hat{d}_{s}^{-1}$ and
\[
[a(\i s), \chi]u= - \p_{j}\chi h_{\i s}^{jk}\p_{k}u - |h_{\i s}|^{-\12}\p_{j}( h_{\i s}^{jk}|h_{\i s}|^{\12}\p_{k}\chi u).
\]
It follows that
\[
\bea
(\chi v| [a_{0}(\i s), \chi]u)_{\cH_{0}^{\pm}}&= (\chi v|  \hat{d}_{s}[a(\i s), \chi]\hat{d}_{s}^{-1}u)_{\cH_{0}^{\pm}}\\
&=-\int_{\Omega^{\pm}}\chi \bar{v}\p_{j}\chi h_{\i s}^{jk}\hat{d}_{s}\p_{k}(\hat{d}_{s}^{-1}u)| h_{0}|^{\12}dsdy\\
&\phantom{=}\, -\int_{\Omega^{\pm}} \hat{d}_{s}|h_{0}|^{\12} |h_{\i s}|^{-\12}\chi \bar{v}\p_{j}(h_{\i s}^{jk}|h_{\i s}|^{\12}\p_{k}\chi \hat{d}_{s}^{-1}u)dsdy\\
&=-\int_{\Omega^{\pm}}\chi \bar{v}\p_{j}\chi h_{\i s}^{jk}\hat{d}_{s}\p_{k}(\hat{d}_{s}^{-1}u)| h_{0}|^{\12}dsdy\\
& \phantom{=}\, +\int_{\Omega^{\pm}} \hat{d}_{s}\p_{j}(\hat{d}_{s}^{-1}\chi \bar{v}) h_{\i s}^{jk}\p_{k}\chi u | h_{0}|^{\12}dsdy,
\eea
\]
where we integrate by parts the second term and use that $\hat{d}_{s}= |h_{0}|^{- 1/4}|h_{\i s}|^{1/4}$. Expanding the derivatives, we obtain:
\[
\bea
&(\chi v| [a_{0}(\i s), \chi]u)_{\cH_{0}^{\pm}}\\
&=-\int_{\Omega^{\pm}}\chi \bar{v}\p_{j}\chi h_{\i s}^{jk}\hat{d}_{s}\p_{k}u| h_{0}|^{\12}dsdy
+\int_{\Omega^{\pm}} \chi\bar{v} \p_{j} \chi h_{\i s}^{jk}\p_{k}(\ln \hat{d}_{s}) u |h_{0}|^{\12}dsdy\\
&\phantom{=}\,+ \int_{\Omega^{\pm}} \chi \p_{j}\bar{v} h_{\i s}^{jk}\p_{k}\chi u  |h_{0}|^{\12}dsdy+\int_{\Omega^{\pm}}\bar{v}\p_{j} \chi h_{\i s}^{jk}\p_{k}\chi u  |h_{0}|^{\12}dsdy\\
&\phantom{=}\,- \int_{\Omega^{\pm} }\chi \bar{v}\p_{j}(\ln(\hat{d}_{s}))h_{\i s}^{jk} \p_{k} \chi u  |h_{0}|^{\12}dsdy.
\eea
\]
 We can use Lemma \ref{lem5.1} {\it iii)}  to bound scalar products involving $h_{\i s}(y)$  by  scalar products involving $h_{0}(y)$. We also know from Lemma \ref{lem5.1} {\it iv)}, {\it v)} that:
 \[
 |\p_{j}(\ln \hat{d}_{s}(y))h_{\i s}^{jk}(y)\p_{k}\chi(y)|\leq c_{0}|\p_{j}\chi(y)h_{0}^{jk}(y)\p_{k}\chi(y)|, \ \ \forall (s, y)\in\Omega.
 \]
 Using  the  Cauchy-Schwarz inequality  we obtain that
 \[
  |(\chi v| [a_{0}(\i s), \chi]u)_{\cH_{0}^{\pm}}|
  \leq c_{0} (\| \chi\|_{\infty}\| \nabla\chi\|_{\infty}+ \| \nabla\chi\|^{2}_{\infty})\| v\|_{H^{1}(\Omega^{\pm})}\| u\|_{H^{1}(\Omega^{\pm})}.
\]
 This completes the proof. \qed
\subsection{The \calde  projectors}
 \begin{definition}\label{def3.1b}
 Let $\Omega\Subset V$ be as in Prop. \ref{p4.1}. The {\em \calde projectors} for $K_{\Omega}$ are the operators
 \[
C_{\Omega}^{\pm}\defeq  \mp \gamma^{\pm}K_{\Omega}^{-1} \gamma^{*}S.
\]
\end{definition}
Note that it is not  a priori clear that $C_{\Omega}^{\pm}$ are well defined, even as maps from $\coinf(\Sigma)^{2}$ to $\cD'(\Sigma)^{2}$. Despite their name, it is even less clear whether $C_{\Omega}^{\pm}$ are projectors on suitable spaces. Let us start by reviewing basic properties of $C_{\Omega}^{\pm}$, which are well known 
if $\Sigma$ is compact. We refer the reader to the book \cite[Chap. 11]{Gr} for details on the compact case.

We recall that the pseudodifferential operator classes $\Psi^{\infty}(\Sigma)$ were introduced in \eqref{eq:pdo}.

\begin{proposition}
 \ben
 \item $C_{\Omega}^{\pm}$ map $\coinf(\Sigma)^{2}$  continuously into $\cinf(\Sigma)^{2}$;
 \item $C_{\Omega}^{\pm}$ are given by $2\times 2$ matrices with entries in $\Psi^{\infty}(\Sigma)$.
 \een
\end{proposition}
\proof  To prove (1) we can replace $K_{\Omega}^{-1}$  by $\varphi_{1}K_{\Omega}^{-1}\varphi_{2}$ for $\varphi_{i}\in \coinf(\Omega)$  equal to $1$ on some neighborhood of $U_{i}\Subset \Sigma$. By Lemma \ref{l4.1} we can then replace $\varphi_{1}K_{\Omega}^{-1}\varphi_{2}$ by  $\varphi_{1}Q\varphi_{2}$ where $Q\in \Psi^{-2}_{\rm c}(\Omega)$ is a properly supported pseudodifferential operator.   The proofs in \cite[Chap. 11]{Gr} can be then applied directly to get (1). 

It also follows  from \cite[Chap. 11]{Gr} that if $\psi\in \coinf(\Sigma)$, then $\psi C_{\Omega}^{\pm}\psi$ is given by a $2\times 2$
 matrix with entries in $\Psi^{\infty}_{{\rm c}}(\Sigma)$.  To prove (2) it hence remains to show that if $\psi_{1}, \psi_{2}\in \coinf(\Sigma)$ have disjoint supports, then $\psi_{1}C_{\Omega}^{\pm}\psi_{2}$ is a smoothing operator on $\Sigma$.  Clearly we can find $\varphi_{i}\in \coinf(\Omega)$ with disjoint supports such that $\varphi_{i}= 1$ near $\{0\}\times \supp \psi_{i}$.  Then in the formula defining $\psi_{1}C_{\Omega}^{\pm}\psi_{2}$  we can replace $K_{\Omega}^{-1}$ by $\varphi_{1}K_{\Omega}^{-1}\varphi_{2}$, which is smoothing by Lemma \ref{l4.1}. Therefore $\psi_{1}C_{\Omega}^{\pm}\psi_{2}$ is smoothing, which completes the proof of (2). \qed

\begin{theoreme}\label{turlututu}  Let $\Omega\Subset V$ be as in Prop. \ref{p4.1}. The following properties hold true:
\ben 
\item one has 
\[
C_{\Omega}^{+}+ C_{\Omega}^{-}= \one\hbox{ on }\coinf(\Sigma)^{2};
\]
\item  setting $\lambda^{\pm}\defeq \pm q \circ C_{\Omega}^{\pm}$ one has
\[
 (f|\lambda^{\pm}f)_{\cS}\geq 0, \ \ \forall f\in\coinf(\Sigma)^{2}.
\]
\een
\end{theoreme}
It follows from the general arguments recalled in \ref{sec1.1.4} that $\lambda^{\pm}$  are a pair of Cauchy surface two-point functions for  the analytic Klein-Gordon operator $P$.
\begin{definition}\label{defdestate}
We denote by $\omega_{\Omega}$ the quasi-free state with Cauchy surface covariances given by the sesquilinear forms $(\cdot | \lambda^{\pm}\cdot)_{\cS}$.
 \end{definition}
 
 Before starting the proof of Thm. \ref{turlututu} we state a result about smooth approximations of the distance function on $(\Sigma, h)$, which follows from \cite[Thm. 1]{AFLR}.
 \begin{proposition}\label{approx}
 Suppose that $\Sigma$ is not compact and let $d(y, y')$ be the geodesic distance for $h_{0}$. Then for any fixed $y^{0}\in \Sigma$, there exists $r\in \cinf(\Sigma)$ such that:
 \[
\begin{array}{rl}
i)& \12 d(y^{0}, y)\leq r(y)\leq 2 d(y^{0}, y),\\[2mm]
ii)& \|\nabla r\|_{\infty}\leq 2.
\end{array}
\]
\end{proposition}
 \noindent {\bf Proof of Thm. \ref{turlututu}.}
For ease of notation we will drop the $\Omega$  subscripts in the sequel. The proof consists of several steps.

{\it Step 1}.
 We first claim that is suffices to prove the theorem for $K_{0}$ instead of $K$. In fact let $C^{\pm}_{0}= \mp \gamma^{\pm}K_{0}^{-1}\gamma_{0}^{*}S_{0}$ be the analogue of $C^{\pm}_{\Omega}$, defined using $K_{0}$, where $\gamma_{0}^{*}$ is  defined as the formal adjoint of $\gamma: \cH_{0}\to \cS$. Note that $\gamma_{0}^{*}= \gamma^{*}$.
 
We recall that:

\[
\begin{array}{l}
K^{-1}_{\Omega}= \hat{d}^{-1}K_{0}^{-1}\hat{d},\ \ \gamma^{(\pm)}= R^{-1}\gamma^{(\pm)}_{0}\hat{d},\\[2mm]
\gamma^{*}= \hat{d}^{-1}\gamma_{0}^{*}(R^{-1})^{*},\ \ S= R^{*}S_{0}R,
\end{array}
\]
which  implies that
\[
C^{\pm}_{\Omega}=  R^{-1}C_{0}^{\pm}R.
\]
Since $q = R^{*}q R$ it suffices to prove the theorem for $C_{0}^{\pm}$.

{\it Step 2}. In Step 2 we  prove (1).

 Let us consider
\[
u^{\pm}= r^{\pm}K_{0}^{-1}\gamma_{0}^{*}S_{0}f, \hbox{ for }f\in \coinf(\Sigma)^{2}.
\]
Using the ellipticity of $K_0$ and the fact that $K_{0}^{-1}$ is a pseudo\-differential operator away from $\p\Omega$, one can prove that $u^\pm\in  \overline{H^{1}_{0}}(\Omega^{\pm})\cap \overline{\cinf}(\Omega^{\pm})$, see Lemma \ref{lem:app1} in the appendix.

Let $f,g\in \coinf(\Sigma)^{2}$. We fix $v\in \coinf(\Omega)$ such that $\gamma v=g$ and set $\tu^{\pm}= \mp r^{\pm}K_{0}^{-1}\gamma^{*}S_{0}f$ so that $\gamma^{\pm}u^{\pm}= C_{0}^{\pm}f$.  We have:
\beq\label{e3.20b}
\bea
(g|S_{0}(\tC^{+}+ \tC^{-})f)_{\cS}&= (\tgam^{+}r^{+}v| S_{0} \tgam^{+}\tu^{+})_{\cS}+ (\tgam^{-}r^{-}v| S_{0} \tgam^{-}\tu^{-})_{\cS}\\[2mm]
&=(r^{+}v| \tK _{0}\tu^{+})_{\tcH_{0}^{+}}-(r^{+}\tK_{0}^{*}v| \tu^{+})_{\tcH_{0}^{+}} \\
&\phantom{= }\, - (r^{-}v| \tK_{0}\tu^{-})_{\tcH_{0}^{-}}+ (r^{-}\tK_{0}^{*}v| \tu^{-})_{\tcH_{0}^{-}}\\[2mm]
&= -(r^{+}\tK_{0}^{*}v| \tu^{+})_{\tcH_{0}^{+}}+ (r^{-}\tK_{0}^{*}v| \tu^{-})_{\tcH_{0}^{-}}\\[2mm]
&=(K_{0}^{*}v| \tK_{0}^{-1}\tgam^{*}S_{0} f)_{\tcH_{0}}= (v|\tgam^{*}S _{0}f)_{\tcH_{0}}\\[2mm]
&= (\tgam v| S_{0} f)_{\cS}= (g| S_{0} f)_{\cS}.
\eea
\eeq
In the second line we used \eqref{alato},  and then we used that $\tK_{0} u^{\pm}=0$ in $\Omega^{\pm}$. Next, in the next lines we used that $v\in \coinf(\Omega)$ and hence  $K_{0}^{*}v= (K_{0})^{*}v$. Since $g$ is arbitrary and $S_{0}$ is injective, \eqref{e3.20b} implies that $\tC_{0}^{+}f+ \tC_{0}^{-}f= f$, which completes the proof of (1).

{\it Step 3}. 
In Step 3  we prove (2).  For simplicity we consider only the case of $C_{0}^{+}$, the case of $C_{0}^{-}$ being similar.
Let $f\in \coinf(\Sigma)^{2}$ and $u^{+}= r^{+} K_{0}^{-1}\gamma_{0}^{*}S_{0}f$.  The idea is to apply the identities \eqref{alato}, \eqref{alatoto} to $v=u=u^{+}$.  However since we do not know if $\gamma^{+}u\in L^{2}(\Sigma)^{2}$, the boundary terms   in these identities may be ill defined. 
Therefore we need some extra approximation argument. This argument
 is superfluous  if $\Sigma$ is compact.

Assume first that $\Sigma$ is non-compact. We fix  $F\in \coinf (\rr)$ equal to $1$ near $0$. We set:
\[
\chi_{n}(y)= F(n^{-1}r(y)), \ \ n\in \nn^{*},
\]
where $r(y)$ is the approximation of the distance function given in Prop. \ref{approx}. Since  $(\Sigma, h_{0})$ is complete we have:
\beq\label{b5.11b}
\begin{array}{rl}
i)&\chi_{n}\in \coinf(\Sigma),\\[2mm]
ii)&\slim_{n\to \infty}\chi_{n}= \one\hbox{ on }\cH_{0}^{\pm},\\[2mm]
iii)&\| \nabla \chi_{n}\|_{\infty}\in O(n^{-1}).
\end{array}
\eeq
If $\Sigma$ is compact we set $\chi_{n}(y)\equiv 1$.

From \eqref{alatoto} we obtain that:
\[
\bea
 &(\chi^{2}_{n}u^{+}| K_{0}u^{+})_{\cH_{0}^{+}}+ (K_{0}u^{+}| \chi_{n}^{2}u^{+})_{\cH^{+}_{0}}\\
 &= 2(\p_{s}\chi_{n}u^{+}| \p_{s}\chi_{n}u^{+})_{\cH^{+}_{0}}+ (u^{+}| (\chi_{n}^{2}a_{0}(\i s)+ a_{0}(\i s)^{*}\chi_{n}^{2})u^{+})_{\cH_{0}^{+}}\\
 &\phantom{=} \,- (\chi_{n}\gamma^{+}u^{+}| q\chi_{n}\gamma^{+}u^{+})_{\cS}.
 \eea
\]
Next, we write:
\[
\bea
&\chi_{n}^{2}a_{0}(\i s)+ a_{0}(\i s)^{*}\chi_{n}^{2}\\
&= \chi_{n}(a_{0}(\i s)+ a_{0}(\i s)^{*})\chi_{n}+ \chi_{n}[\chi_{n}, a_{0}(\i s)]+ [a_{0}(\i s)^{*}, \chi_{n}]\chi_{n},
\eea
\]
which yields:
\begin{equation}
\label{b5.12}
\bea
 &(\chi^{2}_{n}u^{+}| K_{0}u^{+})_{\cH_{0}^{+}}+ (K_{0}u^{+}| \chi_{n}^{2}u^{+})_{\cH^{+}_{0}}\\
 &= 2(\p_{s}\chi_{n}u^{+}| \p_{s}\chi_{n}u^{+})_{\cH^{+}_{0}}+ ( \chi_{n}u^{+}|(a_{0}(\i s)+ a_{0}(\i s)^{*})\chi_{n}u^{+})_{\cH_{0}^{+}}\\
 &\phantom{=} \, +(\chi_{n}u^{+}| [\chi_{n}, a_{0}(\i s)]u^{+})_{\cH^{+}_{0}}+ ([\chi_{n}, a_{0}(\i s)]u^{+}|\chi_{n} u^{+})_{\cH_{0}^{+}}\\
 &\phantom{=} \,  - (\chi_{n}\gamma^{+}u^{+}| q\chi_{n}\gamma^{+}u^{+})_{\cS}.
 \eea
\end{equation}
The first line in \eqref{b5.12} vanishes since $K_{0}u^{+}=0$ in $\Omega^{+}$, the second is positive by \eqref{e4.4}. By Prop. \ref{arlito}  the third line is $O(n^{-1})$.  Since $\gamma^{+}u^{+}= C_{0}^{+}f$ we obtain:
\begin{equation}
\label{b5.13}
(\chi_{n}C_{0}^{+}f| q\chi_{n}C_{0}^{+}f)_{\cS}\geq - Cn^{-1}.
\end{equation}
To complete the proof of (2) we now use the  identity \eqref{alato} combined with reflection in $s$.

Let us set $iu(s, y)\defeq u(-s, y)$ for $u\in \cH_{0}$. Then  since $a_{0}(\i s)^{*}= a_{0}(- \i s)$,  and $\Omega$ is invariant under $i$ we have
\begin{equation}
\label{b5.13t}
i: \Dom K_{0} \xrightarrow{\sim} \Dom K_{0}^{*}, \ \ K_{0}^{*}= i K_{0}i.
\end{equation}
Moreover if  $I = \mat{\one}{0}{0}{-\one}$ one has 
\begin{equation}
\label{b5.13b}
\gamma^{\mp}\circ  i= I \circ \gamma^{\pm}.
\end{equation}

We take $u^{+}= r^{+} K_{0}^{-1}\gamma_{0}^{*}S_{0}f$ as before and $v^{+}= r^{+}i K_{0}^{-1}\gamma_{0}^{*}S_{0}g$ for $g\in \coinf(\Sigma)^{2}$.  Then from \eqref{alato} we obtain
\[
\bea
&(\chi_{n}^{2}v^{+}| K_{0}u^{+})_{\cH_{0}^{+}}- (K_{0}^{*}v^{+}| \chi_{n}^{2} u^{+})_{\cH_{0}^{+}}\\
&=(v^{+} | (\chi_{n}^{2}a_{0}(\i s)- a_{0}(\i s)\chi_{n}^{2})u^{+})_{\cH_{0}^{+}} +(\chi_{n}\gamma^{+}v^{+}| S_{0}\chi_{n}\gamma^{+}u^{+})_{\cS}.
\eea
\]
As above we write:
\[
\chi_{n}^{2}a_{0}(\i s)- a_{0}(\i s)\chi_{n}^{2}= \chi_{n}[\chi_{n}, a_{0}(\i s)]+ [\chi_{n}, a_{0}(\i s)]\chi_{n}, 
\]
which yields
\begin{equation}
\label{b5.14}
\bea
&(\chi_{n}^{2}v^{+}| K_{0}u^{+})_{\cH_{0}^{+}}- (K_{0}^{*}v^{+}| \chi_{n}^{2} u^{+})_{\cH_{0}^{+}}\\
&=(\chi_{n}v^{+} |[\chi_{n}, a_{0}(\i s)]u^{+})_{\cH_{0}^{+}}+ ([a_{0}(\i s)^{*}, \chi_{n}]v^{+}| \chi_{n}u^{+})_{\cH_{0}^{+}}\\
& \phantom{=} \, +(\chi_{n}\gamma^{+}v^{+}| S_{0}\chi_{n}\gamma^{+}u^{+})_{\cS}.
\eea
\end{equation}
 We have $K_{0}u^{+}= 0$ in $\Omega^{+}$ and by \eqref{b5.13t}  $K_{0}^{*}v^{+}=0$ in $\Omega^{+}$ so the first line in  \eqref{b5.14} vanishes. By Prop. \ref{arlito} the second line is $O(n^{-1})$. By \eqref{b5.13b} we see that $\gamma^{+}v^{+}=  I \gamma^{-} K_{0}^{-1}\gamma_{0}^{*}S_{0}g= I C^{-}_{0}g$. Using also that $I^{*} S_{0}= -q$ and $\gamma^{+}u^{+}= - C_{0}^{+}f$,  we obtain that the third line in \eqref{b5.14} equals 
$(\chi_{n}C_{0}^{-}g| q \chi_{n}C_{0}^{+}f)_{\cS}$. Therefore we obtain
\[
(\chi_{n}C_{0}^{-}g| q \chi_{n}C_{0}^{+}f)_{\cS}\in O(n^{-1}).
\]
Taking $g=  f$ this implies that
\begin{equation}
\label{b5.15}
|(\chi_{n}C_{0}^{-}f| q \chi_{n}C_{0}^{+}f)_{\cS}|\in O(n^{-1})
\end{equation}
 We can now complete the proof of (2).  We have for $f\in \coinf(\Sigma)^{2}$ and $n$ large enough
 \[
 \bea
 (f| q C_{0}^{+}f)_{\cS}&= (\chi_{n}^{2}f|q C_{0}^{+}f)_{\cS}= (\chi_{n}f| q \chi_{n}C_{0}^{+}f)_{\cS}\\
 & = (\chi_{n}C_{0}^{+}f| q \chi_{n}C_{0}^{+}f)_{\cS}+ (\chi_{n}C_{0}^{-}f| q \chi_{n}C_{0}^{+}f)_{\cS}\\
 & \geq - Cn^{-1},
 \eea
 \]
 where  we used (1) in the second line and \eqref{b5.13}, \eqref{b5.15} in the third line. Letting $n\to +\infty$ we obtain (2).
 This completes the  proof of the theorem. \qed

\subsection{Purity of $\omega_{\Omega}$}\label{sec4.pure}
 In this subsection we prove that the state $\omega_{\Omega}$ is pure, using the characterization of pure states recalled in Prop. \ref{purity}. 
 
 If $\Sigma$ is compact, this  follows immediately from the fact that $C^{\pm}_{\Omega}$ are projectors (note that $\coinf(\Sigma)= \cinf(\Sigma)$ then so $C^{\pm}_{\Omega}\circ C^{\pm}_{\Omega}$ is then well defined on $\coinf(\Sigma)^{2}$). The fact that $C^{\pm}_{\Omega}$ are projectors in the compact case is well known, see e.g. \cite[Prop. 11.7]{Gr}. 
 
 If $\Sigma$ is not compact, we cannot a priori make sense of $C^{\pm}_{\Omega}\circ C^{\pm}_{\Omega}$, so an extra approximation argument is needed. In the proposition below we assume that $\Sigma$ is not compact and we set $\psi_{n}= \chi_{n}^{2}$, where $\chi_{n}\in \coinf(\Sigma)$ is the sequence of cutoff functions introduced in the proof of Thm. \ref{turlututu}.
\begin{proposition}\label{keyprop}
Let $f\in \coinf(\Sigma)^{2}$. Then:
 \begin{equation}
\label{e.200}
\begin{array}{rl}
i)&\psi_{n}C^{+}_{\Omega}f- C^{+}_{\Omega}\psi_{n}C^{+}_{\Omega}f\to 0,\\[2mm]
ii)&\psi_{n}C^{+}_{\Omega}\psi_{n}C^{+}_{\Omega}f- C^{+}_{\Omega}\psi_{n}C^{+}_{\Omega}\psi_{n}C^{+}_{\Omega}f\to 0,
\end{array}
\end{equation}
in $\cD'(\Sigma)^{2}$ when $n\to \infty$.
\end{proposition}
\proof By the same reasoning as in 
Step 1 in the proof of Thm. \ref{turlututu}, it suffices to prove the analogue of \eqref{e.200} with $C_{\Omega}^+$ replaced by $C_0^+=-\gamma^+ K_{0}^{-1}\gamma^* S_{0}$.

Let us note that the identity \eqref{alato} is of course still valid if $u\in \overline{\coinf}(\Omega^{\pm})$ and $v\in \overline{\cinf}(\Omega^{\pm})$.  It is also valid if $u\in \overline{\coinf}(\Omega^{\pm})$ and $v\in \overline{H^{2}_{\rm loc}}(\Omega^{\pm})$, since all the terms in the identity are still well defined.

For $f\in \coinf(\Sigma)^{2}$ we set
\[
V^{+}_{0}f\defeq - r^{+}K_{0}^{-1}\gamma^{*}S_{0}f.
\]
Let $w\in \overline{\coinf}(\Omega^{+})$ and
\[
v= (K_{0}^{-1})^{*}e^{+}w.
\]
We know that $v\in H^{1}_{0}(\Omega)$ by Prop. \ref{p4.1}, and $v\in H^{2}_{\rm loc}(\Omega)$ using that $e^{+}w\in L^{2}(\Omega)$ and elliptic regularity.  For $u\in\overline{\coinf}(\Omega^{+})$ we obtain from \eqref{alato}:
\begin{equation}
\label{e.20}
(v|K_{0}u)_{\cH_{0}^{+}}- (K_{0}^{*}v|u)_{\cH_{0}^{+}}= (\gamma^{+}v| S_{0}\gamma^{+}u)_{\cS}= (\gamma v| S_{0}\gamma^{+}u)_{\cS},
\end{equation}
where in the last equality we use that   $\gamma^{\pm}v= \gamma v$. In fact  $\gamma v$ is well defined as an element of $H^{3/2}_{\rm loc}(\Sigma)\oplus H^{1/2}_{\rm loc}(\Sigma)$ since $v\in H^{2}_{\rm loc}(\Omega)$. Next, we obtain:
\begin{equation}
\label{e.21}
\bea
(\gamma v| S_{0}\gamma^{+}u)_{\cS}&= (v| \gamma^{*}S_{0}\gamma^{+}u)_{\cH_{0}}= ((K_{0}^{-1})^{*}e^{+}w| \gamma^{*}S_{0}\gamma^{+}u)_{\cH_{0}}\\
&= (e^{+}w| K_{0}^{-1}\gamma^{*}S_{0}\gamma^{+}u)_{\cH_{0}}= - (w| V^{+}_{0}\gamma^{+}u)_{\cH_{0}^{+}}.
\eea
\end{equation}
From \eqref{e.20}, \eqref{e.21} we obtain:
\begin{equation}
\label{e.22}
\bea
(w| u - V^{+}_{0}\gamma^{+}u)_{\cH_{0}^{+}}&= ((K_{0}^{-1})^{*}e^{+}w| K_{0}u)_{\cH_{0}^{+}},\\
&=(v|K_{0}u)_{\cH_{0}^{+}}, \ \ u,w\in \overline{\coinf}(\Omega^{+}).
\eea
\end{equation}

We now fix $f\in \coinf(\Sigma)^{2}$ and $u = V^{+}_{0}f$. By Lemma \ref{lem:app1} in the appendix, we know that $u\in \overline{H^{1}_{0}}(\Omega^{+})\cap \overline{\cinf}(\Omega^{+})$.  We now apply \eqref{e.22} replacing $u$ by $u_{n}= \psi_{n}u$, which belongs to $\overline{\coinf}(\Omega^{+})$.

Since $K_{0}u= 0$ in $\Omega^{+}$, we have $K_{0}\psi_{n}u= [K_{0}, \psi_{n}]u$ and since $\psi_{n}= \chi_{n}^{2}$:
\[
[K_{0}, \psi_{n}]= [a_{0}(\i s), \psi_{n}]= \chi_{n}[a_{0}(\i s), \chi_{n}]+ [a_{0}(\i s), \ch_{n}]\chi_{n}.
\]
By Prop. \ref{arlito} we obtain:
\[
\begin{array}{rl}
|(v| [K_{0}, \psi_{n}]u)_{\cH_{0}^{+}}|&\leq |(\chi_{n}v| [a_{0}(\i s), \chi_{n}]u)_{\cH_{0}^{+}}|+ |([a_{0}^{*}(\i s), \chi_{n}]v| \chi_{n}u)_{\cH_{0}^{+}}|\\[2mm]
&\leq C n^{-1}\|v\|_{H^{1}(\Omega^{+})}\|u\|_{H^{1}(\Omega^{+})}.
\end{array}
\]
Using \eqref{e.22} this yields:
\begin{equation}
\label{e.24}
\bea
|(w| \psi_{n}u- V^{+}_{0}\gamma^{+}\psi_{n}u)_{\cH_{0}^{+}}|&\leq 
Cn^{-1}\| v\|_{H^{1}(\Omega)}\|u\|_{H^{1}(\Omega^{+})}\\
&\leq Cn^{-1}\|e^{+}w\|_{H^{-1}(\Omega)}\|u\|_{H^{1}(\Omega^{+})}.
\eea
\end{equation}
We claim that
 \begin{equation}
\label{arsou}
\|e^{+}w\|_{H^{-1}(\Omega)}\leq C \| w\|_{\overline{H_{0}^{1}}(\Omega^{+})^{*}}.
\end{equation}
Indeed,
for $\tilde{g}\in \coinf(\Omega)$ we have:
\[
\bea
|(e^{+}w| \tilde{g})_{\cH_{0}}|&=|(w|r^{+}\tilde{g})_{\cH_{0}^{+}}|\\
&\leq C\|w\|_{\overline{H_{0}^{1}}(\Omega^{+})^{*}}\|r^{+}\tilde{g}\|_{H^{1}(\Omega^{+})}\\
&\leq C\|w\|_{\overline{H_{0}^{1}}(\Omega^{+})^{*}}\|\tilde{g}\|_{H^{1}(\Omega)},
\eea
\]
which implies \eqref{arsou}. Therefore we deduce from \eqref{e.24} by duality that
if
\[
r_{1,n}\defeq \psi_{n}u- V^{+}_{0}\gamma^{+}\psi_{n}u,
\]
we have
\begin{equation}
\label{e.25}
\|r_{1, n}\|_{H^{1}(\Omega)}\leq C n^{-1}\| u\|_{H^{1}(\Omega^{+})}.
\end{equation}
Hence, $r_{1, n} \to 0$ in $\overline{H^{1}}(\Omega^{+})$ as $n\to\infty$. 

We now apply \eqref{e.22} once again replacing $u$ by $v_{n}= \psi_{n}V^{+}_{0}\gamma^{+}\psi_{n}u$. We obtain since $K_{0}V^{+}_{0}\gamma^{+}\psi_{n}u=0$ in $\Omega^{+}$:
\[
(w| v_{n}- V^{+}_{0}\gamma^{+}v_{n})_{\cH_{0}^{+}}= (v| [K_{0}, \psi_{n}]V^{+}_{0}\gamma^{+}\psi_{n}u)_{\cH_{0}^{+}},
\]
hence
\[
|(w| v_{n}- V^{+}_{0}\gamma^{+}v_{n})_{\cH_{0}^{+}}|\leq Cn^{-1}\| e^{+}w\|_{H^{-1}(\Omega)}\| V^{+}_{0}\gamma^{+}\psi_{n}u\|_{H^{1}(\Omega^{+})}. 
\]
By \eqref{e.25} we have
\[
\| V^{+}_{0}\gamma^{+}\psi_{n}u\|_{H^{1}(\Omega^{+})}\leq C\|\psi_{n}u\|_{H^{1}(\Omega^{+})}+ Cn^{-1}\| u\|_{H^{1}(\Omega^{+})}\leq C \|u\|_{H^{1}(\Omega^{+})},
\]
since $\|\nabla \psi_{n}\|_{\infty}\leq C$.  Finally we obtain as before that:
\begin{equation}
\label{e.26}
r_{2, n}\defeq \psi_{n}V^{+}_{0}\gamma^{+}\psi_{n}u- V^{+}_{0}\gamma^{+}\psi_{n}V^{+}_{0}\gamma^{+}\psi_{n}u\to 0\hbox{ in }\overline{H^{1}}(\Omega^{+}),
\end{equation}
as $n\to \infty$.

We note then that if $U\Subset \Omega$ and $U^{+}= U\cap \{s>0\}$, we have  $K_{0}\psi_{n}u=0$ in $U^{+}$ for $n\gg 1$, hence $K_{0}r_{1, n}= K_{0}r_{2, n}=0$ in $U^{+}$ for $n\gg 1$. We can now use the argument of `partial hypoellipticity at the boundary' (see e.g. \cite[page 311]{Gr}), which we now briefly explain.

We can introduce local coordinates on $\Sigma$ near $y^{0}\in \Sigma$, and map $U$ to a neighborhood $V$ of $(0, 0)$ in $\rr^{1+d}$ for $d= \dim\Sigma$. We denoting by $H^{m, \ell}(\rr^{1+d})$ the space of $u\in \cS'(\rr^{1+d})$ such that $\langle D_{x}\rangle^{m}\langle D_{y}\rangle^{\ell}u\in L^{2}(\rr^{1+d})$, and then using the coordinates we define the spaces  ${H^{1+k, -k}_{\rm loc}}(\Omega)$ and $\overline{H^{1+k, -k}_{\rm loc}}(\Omega^{+})$ (the definition depends on the choice of coordinates, but this is not important here). Then one deduces, using that $K_{0}r_{i ,n}=0$ in $U^{+}$ that $r_{i, n}\to 0$ also in  $\overline{H^{1+k, -k}_{\rm loc}}(\Omega^{+})$ for any $k\in \nn$.

We can now safely apply  the boundary value operator $\gamma^{+}$ and deduce from \eqref{e.25}, \eqref{e.26} that:
\[
\gamma^{+}r_{i, n}\to 0\hbox{ in }\cD'(\Sigma)^{2}.
\]
Using the definitions of $u$, $r_{i, n}$  and the fact that $C^{+}_{0}= \gamma^{+}V^{+}_{0}$ this yields:
\[
\begin{array}{l}
\psi_{n}C^{+}_{0}f- C^{+}_{0}\psi_{n}C^{+}_{0}f\to 0,\\[2mm]
\psi_{n}C^{+}_{0}\psi_{n}C^{+}_{0}f- C^{+}_{0}\psi_{n}C^{+}_{0}\psi_{n}C^{+}_{0}f\to 0,
\end{array}
\]
in $\cD'(\Sigma)^{2}$ when $n\to \infty$, which entails the desired result. \qed

\begin{proposition}\label{puro}
 The state $\omega_{\Omega}$ is pure.
\end{proposition}
\proof 
By Prop. \ref{purity}  it suffices to find, for each $f\in \coinf(\Sigma)^{2}$, a sequence $f_{n}\in\coinf(\Sigma)^{2}$ such that:
\beq\label{e.27}
\lim_{n\to +\infty}\frac{| \bar{f}\cdot q f_{n}|^{2}}{ \bar{f}_{n}\cdot(\lambda^{+}+ \lambda^{-})f_{n}}= \bar{f}\cdot(\lambda^{+}+ \lambda^{-})f.
\eeq
We take
\[
f_{n}= \psi_{n}(C^{+}_{\Omega}- C^{-}_{\Omega})f,
\]
and note first that
\beq\label{e.28}
\bar{f}\cdot q f_{n}= \bar{f}\cdot q \psi_{n}(C^{+}_{\Omega}- C^{-}_{\Omega})f= \overline{\psi_{n}f}\cdot (\lambda^{+}+ \lambda^{-})f= \overline{f}\cdot (\lambda^{+}+ \lambda^{-})f
\eeq
for $n\gg 1$, since $f$ has compact support. It remains to compute the limit of the denominator in \eqref{e.27}, which we will denote by $I_{n}$ in the sequel. Since $C^{+}_{\Omega}- C^{-}_{\Omega}= 2C^{+}_{\Omega}-1$ and $\lambda^{+}+ \lambda^{-}= q(2C^{+}_{\Omega}-1)$, we obtain:
\[
I_{n}= \overline{ \psi_{n}(2C^{+}_{\Omega}-1)f}\cdot q (2C^{+}_{\Omega}-1)\psi_{n}(2C^{+}_{\Omega}-1)f.
\]
Next  
\[
(2C^{+}_{\Omega}-1)\psi_{n}(2C^{+}_{\Omega}-1)= 4 C^{+}_{\Omega}\psi_{n}C^{+}_{\Omega}- 2 C^{+}_{\Omega}\psi_{n}- 2 \psi_{n}C^{+}_{\Omega}+ \psi_{n},
\]
 which yields:
 \[
\bea
I_{n}&= 8 \overline{ \psi_{n}C^{+}_{\Omega}f}\cdot q C^{+}_{\Omega}\psi_{n}C^{+}_{\Omega}f- 4  \overline{ \psi_{n}C^{+}_{\Omega}f}\cdot q C^{+}_{\Omega}\psi_{n}f\\
&\phantom{=}\, - 4\overline{ \psi_{n}C^{+}_{\Omega}f}\cdot q\psi_{n}C^{+}_{\Omega}f+ 2 \overline{ \psi_{n}C^{+}_{\Omega}f}\cdot q \psi_{n}f\\
&\phantom{=}\, - 4 \overline{\psi_{n}f}\cdot q  C^{+}_{\Omega}\psi_{n}C^{+}_{\Omega}f+ 2  \overline{\psi_{n}f}\cdot q  C^{+}_{\Omega}\psi_{n}f\\
&\phantom{=}\, + 2 \overline{\psi_{n}f}\cdot q \psi_{n}C^{+}_{\Omega}f -  \overline{\psi_{n}f}\cdot q \psi_{n}f.
\eea
\]
Using that $q C^{+}_{\Omega}$ is hermitian  by Thm. \ref{turlututu} we obtain:
 \[
\bea
I_{n}&= 8 \overline{f}\cdot q C^{+}_{\Omega}\psi_{n}C^{+}_{\Omega}\psi_{n}C^{+}_{\Omega}f- 4  \overline{ f}\cdot q C_{\Omega}^{+}\psi_{n}C_{\Omega}^{+}\psi_{n}f\\
&\phantom{=}\, - 4\overline{ f}\cdot qC^{+}_{\Omega}\psi^{2}_{n}C^{+}_{\Omega}f+ 2 \overline{f}\cdot q C^{+}_{\Omega}\psi_{n}^{2}f\\
&\phantom{=}\, - 4 \overline{f}\cdot q \psi_{n} C^{+}_{\Omega}\psi_{n}C^{+}_{\Omega}f+ 2  \overline{f}\cdot q  \psi_{n}C^{+}_{\Omega}\psi_{n}f\\
&\phantom{=}\, + 2 \overline{f}\cdot q \psi_{n}^{2}C^{+}_{\Omega}f -  \overline{f}\cdot q \psi^{2}_{n}f.
\eea
\]
Next we use that $f$ has compact support, hence for $n\gg 1$ we have:
 \[
\bea
I_{n}&= 8 \overline{f}\cdot q C^{+}_{\Omega}\psi_{n}C^{+}_{\Omega}\psi_{n}C^{+}_{\Omega}f- 4  \overline{ f}\cdot q C^{+}_{\Omega}\psi_{n}C^{+}_{\Omega}f\\
&\phantom{=}\, - 4\overline{ f}\cdot qC^{+}_{\Omega}\psi^{2}_{n}C^{+}_{\Omega}f+ 2 \overline{f}\cdot q C^{+}_{\Omega}f\\
&\phantom{=}\, - 4 \overline{f}\cdot q C^{+}_{\Omega}\psi_{n}C^{+}_{\Omega}f+ 2  \overline{f}\cdot q C^{+}_{\Omega}f\\
&\phantom{=}\, + 2 \overline{f}\cdot q C^{+}_{\Omega}f -  \overline{f}\cdot q f.
\eea
\]
We now apply  Prop. \ref{keyprop} {\it ii)} and obtain that:
 \[
\bea
I_{n}&= 8\overline{f}\cdot q \psi_{n}C^{+}_{\Omega}\psi_{n}C^{+}_{\Omega}f- 4  \overline{ f}\cdot q \psi_{n}C^{+}_{\Omega}f\\
&\phantom{=}\, - 4\overline{ f}\cdot q\psi^{2}_{n}C^{+}_{\Omega}f+ 2 \overline{f}\cdot q C^{+}_{\Omega}f\\
&\phantom{=}\, - 4 \overline{f}\cdot q \psi_{n}C^{+}_{\Omega}f+ 2  \overline{f}\cdot q C^{+}_{\Omega}f\\
&\phantom{=}\, + 2 \overline{f}\cdot q C^{+}_{\Omega}f -  \overline{f}\cdot q f+ o(1).
\eea
\]
Using again that $f$ has  compact support we obtain:
\[
\bea
I_{n}&= 8\overline{f}\cdot q C^{+}_{\Omega}\psi_{n}C^{+}_{\Omega}f- 8  \overline{ f}\cdot qC^{+}_{\Omega}f\\
&\phantom{=}\, + 2  \overline{f}\cdot q C^{+}_{\Omega}f -  \overline{f}\cdot q f+ o(1)\\
&= \overline{f}\cdot q(2C^{+}_{\Omega}- 1)f+ o(1).
\eea
\]
Therefore 
\[
\lim_{n\to \infty} \bar{f}_{n}\cdot (\lambda^{+}+ \lambda^{-})f_{n}=\bar{f}\cdot (\lambda^{+}+ \lambda^{-})f,
\]
which using \eqref{e.28} completes the proof of \eqref{e.27}. \qed

\subsection{An example} For illustration let us consider the special case of an ultra-static metric
\[
g=-dt^2+h(y) dy^2,
\]
where $(\Sigma,h(y) dy^2)$ is complete, and a constant mass term $V(x)= m^{2}\geq 0$. Then,
\[
P=  \p_{t}^{2}+ a, \ \ K=-\p_{s}^{2}+ a, 
\]
where $a= - \Delta_{h}+ m^{2}$ is essentially self-adjoint in $L^2(\Sigma,|h|^{\12}dy)$. We denote by $\epsilon\geq 0$ the square root of the closure of $a$. 

As the set $\Omega$ we can take $]-T, T[\times \Sigma$ for  $T>0$. The closed operator $K_\Omega$ corresponds to Dirichlet boundary conditions at $s= \pm T$. 

Let us denote by $\theta(s)$ the Heaviside step function.

We can easily compute $K_{\Omega}^{-1}$, namely $K_{\Omega}^{-1}v= u -r$ where
\[
u(s)= (2\epsilon)^{-1}\int_{-\infty}^{+\infty} \theta(s-s')\e^{-(s-s')\epsilon}+ \theta(s'-s)\e^{(s-s')\epsilon}v(s')ds',
\]
and
\[
r(s)= (2\epsilon)^{-1}(\e^{ 4T\epsilon}-1)^{-1}(\e^{(2T-s)\epsilon}v^{+}- \e^{s\epsilon}v^{+}- \e^{-s\epsilon}v^{-} + \e^{(s+2T)\epsilon}v^{-}),
\]
and
\[
v^{\pm}= \int_{-T}^{T}\e^{\pm s'}v(s')ds'.
\]
A simple computation shows that the \calde projector  is
\beq\label{eq:ererg}
C_{\Omega}^{+}= \12\mat{1}{\epsilon^{-1}\th (T\epsilon)}{\epsilon \coth(T\epsilon)}{1}.
\eeq
Note that the infrared singularity (occurring if $0\notin \rs(\epsilon)$) is completely `smoothed out' by the Dirichlet boundary condition at $\{s=-T\}\cup\{s=T\}$.

Suppose now for simplicity that $\epsilon>0$. In the limit when $T\to\infty$, the right hand side of \eqref{eq:ererg} converges to a projection that corresponds to the \emph{ground state} for $P$.

 \section{Analytic Hadamard property}\label{sec5}\init
 In this section we prove that the  quasi-free state $\omega_{\Omega}$ constructed in Thm. \ref{turlututu} is an analytic Hadamard state. We  first recall well-known facts about the representation of distributions as sums of boundary values of holomorphic functions. We refer the reader to \cite[Chap. 3]{H2}, \cite [Sec. 3.4]{Kom} for details.

\subsection{Distributions as boundary values of holomorphic functions}\label{sec5.1}
\subsubsection{Notation}\label{sec5.1.1} We first introduce some notation.

\begin{notations}
\item In  the sequel a cone of vertex $0$ in $\rr^{n}$ which is convex, open and proper will be  simply called a {\em convex open cone}. If $\Gamma, \Gamma'$ are two cones of vertex $0$ in $\rr^{n}$ we write $\Gamma'\Subset \Gamma$ if $(\Gamma'\cap\mathbb{S}^{n-1})\Subset (\Gamma\cap \mathbb{S}^{n-1})$.

\item If $\Gamma$ is a  convex open cone we denote by
\[
\Gamma^{o}\defeq \{\xi\in \rr^{n}: \ \xi\cdot y\geq 0, \ \ \forall y\in \Gamma\}
\]
its polar.  $\Gamma^{o}$ is a closed convex cone.

\item  Let $\Omega\subset \rr^{n}$ be open and let $\Gamma\subset \rr^{n}$ be a convex open cone. Then a domain $D\subset \cc^{n}$ is called a {\em tuboid of profile} $\Omega+ \i \Gamma$ if:
\ben
\item $D\subset \Omega+ \i \Gamma$,
\item for any $x_{0}\in \Omega$ and  any  subcone $\Gamma'\Subset \Gamma$ there exists  a neighborhood $\Omega'$ of $x_{0}$ in $\Omega$ and $r>0$ such that 
\[
\Omega'+ \i \{y\in \Gamma': 0<|y|\leq r\}\subset D.
\]
\een
\item  If $D\subset \cc^{n}$ is open, we denote by $\mo(D)$ the space of holomorphic functions in $D$.
\item We write $F\in \mo(\Omega+\i \Gamma 0)$ if $F\in \mo(D)$ for some tuboid $D$ of profile $\Omega+ \i \Gamma$. 
\item If  $F\in \mo(D)$ for some tuboid $D$ of profile $\Omega+ \i \Gamma$, we write  $F\in \mo_{\rm temp}(\Omega+ \i \Gamma 0)$  and say that $F$ is {\em temperate}, if for any $K\Subset \Omega$, any  subcone $\Gamma'\Subset \Gamma$ and $r>0 $ such that $K+\i\{y\in  \Gamma': 0<|y|\leq r\}\subset D$, there exists $C, r>0, N\in \nn$ such that
\begin{equation}
\label{e5.-2}
|F(x+ \i y)|\leq C |y|^{-N}, \ \ x\in K, \  y\in \Gamma',0<|y|\leq r.
\end{equation}
\end{notations}

\subsubsection{Boundary values of holomorphic functions}\label{sec5.1.2}

If $F\in \mo_{\rm temp}(\Omega+\i \Gamma 0)$  the limit 
\beq\label{e5.-1}
\lim_{\Gamma'\ni y\to 0}F(x+ \i y)=f(x)\hbox{ exists in }\cD'(\Omega),
\eeq
for any  $\Gamma'\Subset \Gamma$ and is denoted  by $F(x+ \i \Gamma 0)$,
 (see e.g. \cite[Thm. 3.6]{Kom}).
 
If $\Gamma_{1}, \dots ,\Gamma_{N}$ are convex open cones such that $\bigcup_{1}^{N}\Gamma_{i}^{o}= \rr^{n}$ then any $u\in \cD'(\Omega)$ can be written as
\beq\label{e5.-3}
u(x)= \sum_{j=1}^{N}F_{j}(x+ \i \Gamma_{j}0), 
\eeq
for some $F_{j}\in \mo_{\rm temp}(\Omega+ \i \Gamma_{j}0)$.  The non-uniqueness of the decomposition \eqref{e5.-3} is described by {\em Martineau's edge of the wedge theorem}, which states that
\[
\sum_{j=1}^{N}F_{j}(x+ \i \Gamma_{j}0)=0 \hbox{ in }\cD'(\Omega)
\]
for $F_{j}\in \mo_{\rm temp}(\Omega+ \i \Gamma_{j}0)$ iff there exist $H_{jk}\in \mo_{\rm temp}(\Omega+ \i \Gamma_{jk}0)$ for $\Gamma_{jk}= (\Gamma_{j}+ \Gamma_{k})^{\rm conv}$ ($A^{\rm conv}$ denotes the convex hull of $A$) such that
\[
F_{j}= \sum_{k}H_{jk} \hbox{ in } \Omega+ \i\Gamma_{j}, \ \ H_{jk}= - H_{kj} \hbox{ in } \Gamma_{jk},
\]
 see for example \cite[Thm. 3.9]{Kom}.

\subsection{The analytic \wavefront set}\label{sec5.2}
We now recall the definition of the {\em analytic \wavefront set} of a distribution on $\rr^{n}$.

\begin{definition}\label{def:wfa}
Let $u\in \cD'(\Omega)$ for $\Omega\subset \rr^{n}$ open and $(x^{0}, \xi^{0})\in \Omega\times \rr^{n}\backslash\{0\}$. Then $(x^{0}, \xi^{0})$ does not belong to the {\em analytic \wavefront set} $\WFA(u)$ if there exists $N\in\nn$, a neighborhood $\Omega'$ of $x^{0}$ in $\Omega$ and  convex open cones $\Gamma_{j}$, $1\leq j\leq N$, such that
\[
u(x)= \sum_{j=1}^{N}F_{j}(x+ \i \Gamma_{j}0) \hbox{ over }\Omega',
\]
for $F_{j}\in \mo_{\rm temp}(\Omega'+ \i \Gamma_{j}0)$, $1\leq j\leq N$, and $F_{j}$ holomorphic near $x^{0}$ if $\xi^{0}\in \Gamma_{j}^{o}$. 
\end{definition}
The equivalence of the above definition of the analytic \wavefront set with other ones, in particular with the one introduced by H\"{o}rmander (see e.g. \cite[Def. 8.4.3]{H2}, has been proved by Bony in \cite{B}. 

 The analytic \wavefront set is covariant under analytic diffeomorphisms, which allows to extend its definition to distributions on a real analytic manifold in the usual way.

 We will often work with  open sets of the form $\Omega= I\times Y$ for $I\subset \rr$ an open interval and $Y\subset \rr^{n-1}$ open, writing $x\in \Omega$ as $(t, y)$.    If $\Gamma= ]0, +\infty[$, we will write simply $I\pm \i 0$ for the profiles $I\pm \i \Gamma 0$. We denote by $\mo_{\rm temp}(I\pm \i 0; \cD'(Y))$ the space of temperate $\cD'(Y)$-valued holomorphic functions on some tuboid $D$ of profile  $I\pm \i 0$. This means that for each  $K\Subset I$  there exist $r>0, N\in \nn$ such that  for each  bounded set $B\subset \cD(Y)$ there exist $C_{B}>0$ such that
 \[
 \sup_{\varphi\in B}|\langle u(z, \cdot), \varphi(\cdot)\rangle_{Y}|\leq C_{B}|\Im z|^{-N}, \ \ \Re z\in K, \pm \Im z >0, |\Im z|\leq r, 
 \]
  where $\langle \cdot, \cdot \rangle_{Y}$ is the duality bracket between $\cD'(Y)$ and $\cD(Y)$.

Let us set $\varphi_{z}(s)= (s-z)^{-1}$ for $z\in \cc\backslash \rr$.
If $u\in \cD'(\rr\times \rr^{n-1})$ has compact support, then   \[
F(z, y)= \frac{1}{2\i \pi}\langle \varphi_{z}(\cdot), F(\cdot, y)\rangle_{\rr}
\]
belongs to $\mo_{\rm temp}(\rr\pm \i 0; \cD'(\rr^{n-1}))$ and 
\[
u(s, y)= F(s+ \i 0, y)- F(s- \i 0, y),
\]
where $F(s\pm \i 0, y)= \lim_{\epsilon\to 0^{+}}F(s+ \i 0, y)$ in $\cD'(\rr\times \rr^{n-1})$. 

\subsection{Proof of  the analytic Hadamard property}\label{sec5.3}

Let  us fix   a  neighborhood $U$ of $\{0\}\times \Sigma$ in $\rr\times \Sigma$,  a neighborhood $U'$ of $\Sigma$ in $M$ and 
\[
\chi: (U, - dt^{2}+ h_{t}(y)dy^{2})\to (U', g)
\]
the analytic isometric diffeomorphism given by the Gaussian normal coordinates to $\Sigma$ in $(M,g)$.  We recall that $\chi^{*}P= \p_{t}^{2}+ r(t, y)\p_{t}+ a(t, y, \p_{y})=P(t, y, \p_{t}, \p_{y})$.  Since $P$
has analytic coefficients, $P$ extends  holomorphically in $z= t+ \i s$ to a neighborhood $W$ of $\{0\}\times \Sigma$ in $\cc\times \Sigma$.  This holomorphic extension will be denoted by 
\[
P_{z}= P(z, y, \p_{z}, \p_{y}),
\]
and $P(t, y, \p_{t}, \p_{y})$ will be denoted by $P_{t}$.  The Wick rotated operator $K= P(\i s, y, - \i \p_{s}, \p_{y})$,  defined on a neighborhood $V$ of $\{0\}\times \Sigma$  in $\i \rr\times \Sigma$, will henceforth be denoted by $P_{\i s}$.

We also fix an open set $\Omega\subset V$ as in Prop. \ref{p4.1} and recall that  $C^{\pm}_{\Omega}$  are the associated \calde projectors constructed in Sect. \ref{seccald}.  Moreover for $f\in \cD'(\Sigma)^{2}$  we denote by  $U_{\Sigma}f\in \cD'(U)$ the solution  of the  Cauchy problem:
 \[
\begin{cases}
P_{t}u=0\hbox{ in }U\\ \rho_{\Sigma} u=f.
\end{cases}
\]
We will first prove in Prop. \ref{propito} a result of independent interest about a key property  of $C_{\Omega}^{\pm}$.
In the $C^{\infty}$ case (see \cite[Thm. 3.12]{GW2}) it is known that this property implies that  $\omega$ is a $\cinf$ Hadamard state. 

 We are indebted to Pierre Schapira for  crucial help with the proof of Prop. \ref{prop5.1}, who also gave an  extension of this result to the framework of $\mathcal{D}-$modules in \cite{Sch}.
\begin{proposition}\label{prop5.1}
 We have:
 \[
\WFA (U_{\Sigma}C_{\Omega}^{\pm}f)\subset \cN^{\pm} \ \  \forall f\in\cE'(\Sigma)^{2}.
\]
\end{proposition}
\proof
We only prove the proposition for $C_{\Omega}^{+}$, the proof for $C_{\Omega}^{-}$ being similar.

In the sequel we set $I= ]-\delta, \delta[$ where $\delta>0$ will be chosen small enough.  We set 
\[
v\defeq -K_{\Omega}^{-1}\gamma^{*}Sf, \ \ g\defeq \gamma^{+}v= C_{\Omega}^{+}f, \ \ u\defeq U_{\Sigma}C_{\Omega}^{+}f.
\]
The proof will be split in several steps.
 
 {\it Step 1.}

By the analytic propagation of singularities theorem (see \cite[Thm. 3.3']{kawai} or \cite[Thm. 7.1]{H4}), it suffices to prove that  $\WFA( u)\subset \cN^{+}$ over $I\times Y_{1}$    for a neighborhood $Y_{1}\Subset \Sigma$  of some $y^{0}\in \Sigma$. By a partition of unity argument, we can also assume that $\supp f\Subset Y$  where  $Y$ is an arbitrary open neighborhood of $Y_{1}$. In fact  since $P_{\i s}v= \gamma^{*}Sf$ and $P_{\i s}$  is an elliptic operator with analytic coefficients, we deduce from 
the Morrey-Nirenberg theorem (see e.g. \cite[Thm. 7.5.1]{H1}) that if $f= 0$ near $Y_{1}$, then $v$ is analytic near $\{0\}\times Y_{1}$ hence $g$ is analytic near $Y_{1}$. This implies that  $u$ is analytic near $y^{0}$.

Another observation is that by finite propagation speed,  if $g_{1}=g$ near $Y_{1}$, then $U_{\Sigma}g= U_{\Sigma}g_{1}$ near  $I\times Y_{1}$.  Therefore we can fix cutoff functions $\psi\in \coinf(I), \tilde{\psi}\in \coinf(Y)$ with $\psi= 1$ near $0$, $\tilde{\psi}= 1$ near $Y_{1}$ and  replace $v$ by $\chi v$ for $\chi(s, y)= \psi(s)\tilde{\psi}(y)$ so that 
\beq\label{e5.0ER}
\gamma^{+}\chi v= \tilde{\psi}g.
\eeq
{\it Step 2.}

 Writing $z= t+ \i s$ motivates the following notation that we will use in the sequel:
\[
\begin{array}{l}
I^{\ri/\le}= I\cap \{\pm t>0\}, \ \ I^{\pm}= I\cap \{\pm s>0\}, \\[2mm]
D= I \times \i I, \ \ D^{\pm}= I\times  \i I^{\pm}, \ \ D^{\ri/\le}= I^{\ri/\le}\times  \i I.
\end{array}
\]
By Subsect. \ref{sec5.1} we can write
\beq\label{e5.-10}
\chi v(s, y)= v^{\ri}(\i s + 0,y)- v^{\le}(\i s - 0,y),
\eeq
where $v^{\ri/\le}$ are the restrictions to $D^{\ri/\le}\times Y_{1}$ of
\[
F(z, y)= -\frac{1}{2\i \pi}\langle\varphi_{- \i z}(\cdot), \chi v(\cdot, y)\rangle_{\rr}.
\]
Since $P_{\i s}v= \gamma^{*}Sf= \delta(s)\otimes h_{0}(y)+ \delta'(s)\otimes h_{1}(y)$ for $h_{0}, h_{1}\in \cE'(\Sigma)$, we have
\[
P_{\i s }\chi v=  \delta(s)\otimes h_{0}(y)+ \delta'(s)\otimes h_{1}(y)\hbox{ on } I\times Y_{1}.
\]
Using that $\delta(s)= \frac{1}{2\i \pi}(\frac{1}{s+ \i 0}- \frac{1}{s- \i 0})$, this implies   that
\[
P_{z}v^{\ri/\le}= w\hbox{ in } D^{\ri/\le}\times Y_{1},
\]
where
\[
w(z, y)= \frac{1}{2\pi z}\otimes h_{0}(y)+\frac{1}{2\i \pi z^{2}}\otimes h_{1}(y)+ r(z, y),
\]
and  $r(z, y)\in \mo(D; \cD'(N_{1}))$.  Note that $w\in \mo_{\rm temp}(D^{+}; \cD'(Y_{1}))$.
Let us now set (see also Fig. 1):
\[
u^{\ri/\le}(t, y)= v^{\ri/\le}(t+ \i 0, y)_{| I^{\ri/\le}}\in \cD'(I^{\ri/\le}\times Y_{1}).
\]
We have
\[
P_{t}u^{\ri/\le}(t, y)= P_{z}v^{\ri/\le}(t+ \i 0, y)= w(t+ \i 0, y) \hbox{ in  }I^{\ri/\le}\times Y_{1}.
\]
 \begin{figure}[H]\label{fig1}
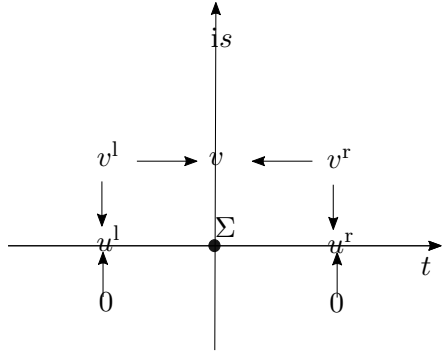\caption{Relation between $v$, $v^{\ri/\le}$ and $u^{\ri/\le}$.}
\end{figure}
Since $P_{t}$ is hyperbolic with respect to $dt$, we can extend $u^{\ri/\le}$ to $\tilde{u}^{\ri/\le}\in \cD'(I\times Y_{1})$ which solve
\[
\begin{cases}
P_{t}\tilde{u}^{\ri/\le}= w(t+ \i 0, y) \hbox{ in }I\times Y_{2},\\
 \tilde{u}^{\ri/\le}(t, y)= u^{\ri/\le}(t, y) \hbox{ in } I^{\ri/\le}\times Y_{2},
\end{cases}
\]
where $Y_{2}\Subset Y_{1}$ is a neighborhood of $y^{0}$ in $\Sigma$.   Since $w(t+ \i0, y )$ and $u^{\ri/\le}(t, y )$ are boundary values of holomorphic functions from  $D^{+}$, we know (see for example \cite[Thm. 4.3.10]{Ka}) that
\[
\begin{array}{rl}
\WFA\left( w(t+ \i 0, y)\right)\subset \{\tau\geq 0\}\hbox{ over }I\times Y_{1},\\[2mm]
\WFA \big( u^{\ri/\le}(t, y)\big)\subset \{\tau\geq 0\}\hbox{ over }I^{\ri/\le}\times Y_{1}.
\end{array}
\]
By  propagation of singularities (see \cite[Thm. 3.3']{kawai} or \cite[Thm. 7.1]{H4}),  this implies that
\[
\WFA\big( \tilde{u}^{\ri/\le}(t, y)\big)\subset  \{\tau\geq 0\}\hbox{ over }I\times Y_{2}.
\]
Let us denote by $\alpha= (\alpha_{1}, \dots, \alpha_{n-1})$ or $\beta= (\beta_{1}, \dots, \beta_{n-1})$ the elements of $\{-1, 1\}^{n-1}=A=B$ and by $\gamma=(\gamma_{1}, \dots, \gamma_{n-1})$ the elements of $C= A\sqcup B$, where $\sqcup$ denotes the disjoint union. 

 We set $\Delta_{\gamma}= \{y\in \rr^{n-1}: y_{j}\gamma_{j}>0\}$, and $\Gamma_{\gamma}= ]0, +\infty[\times \Delta_{\gamma}$. 
Since the polar cones $\Gamma_{\gamma}^{o}$  cover $\WFA(\tilde{u}^{\ri/\le})$ over $I\times Y_{2}$, we can by \cite[Thm. 3.9]{Kom} write  $\tilde{u}^{\ri/\le}$ as
\beq\label{papepresident.e1}
\tilde{u}^{\ri/\le}(x)= \sum_{\alpha\in A} U^{\ri/\le}_{\alpha}(x+ \i \Gamma_{\alpha}0),\hbox{ over }I\times Y_{2},
\eeq
for $U^{\ri/\le}_{\alpha}\in \mo_{\rm temp}((I\times Y_{2})+ \i \Gamma_{\alpha}0)$.

Similarly we have
\beq
\label{papepresident.e2}
u^{\ri/\le}(x)= v^{\ri/\le}(t+ \i 0, y)= \sum_{\beta\in B}V_{\beta}^{\ri/\le}(x+ \i \Gamma_{\beta}0)\hbox{ over }I^{\ri/\le}\times Y_{2},
\eeq
for $V^{\ri/\le}_{\beta}\in \mo_{\rm temp}((I^{\ri/\le}\times Y_{2})+ \i \Gamma_{\beta}0)$. By Martineau's edge of the wedge theorem, see \cite[Thm. 3.9]{Kom}, we can find $H^{\ri/\le}_{\gamma,\gamma'}\in \mo_{\rm temp}((I^{\ri/\le}\times Y_{2})+ \i \Gamma_{\gamma,\gamma'}0)$, for $\Gamma_{\gamma,\gamma'}= (\Gamma_{\gamma}+ \Gamma_{\gamma'})^{\rm conv}$ such that $H_{\gamma', \gamma}= - H_{\gamma, \gamma'}$ and:
\beq\label{papepresident.e3}
U^{\ri/\le}_{\alpha}= \sum_{\gamma'\in C}H^{\ri/\le}_{\alpha, \gamma'}\hbox{ on }(I^{\ri/\le}\times Y_{2})+ \i \Gamma_{\alpha}0,
\eeq
\beq\label{papepresident.e4}
V^{\ri/\le}_{\beta}=-\sum_{\gamma'\in C}H^{\ri/\le}_{\beta,\gamma'}\hbox{ on }(I^{\ri/\le}\times Y_{2})+ \i \Gamma_{\beta}0.
\eeq
Let us set
\[
\tilde{v}^{\ri/\le}(z, y)= \sum_{\alpha\in A}U_{\alpha}^{\ri/\le}(z, y+ \i \Delta_{\alpha}0)\in \mo_{\rm temp}(D^{+}; \cD'(Y_{2})),
\]
so that $\tilde{u}^{\ri/\le}(t, y)= \tilde{v}^{\ri/\le}(t+ \i 0, y)$
by \eqref{papepresident.e1}. We obtain that
\[
\bea
\tilde{v}^{\ri/\le}(z, y)&=\sum_{\alpha\in A, \gamma'\in C}H^{\ri/\le}_{\alpha,\gamma'}(z, y+ \i\Delta_{\alpha}0)\\[2mm]
&=-\sum_{\beta'\in B, \alpha\in A}H^{\ri/\le}_{\beta', \alpha}(z,y+\i \Delta_{\alpha}0)\\[2mm]
&=-\sum_{\beta'\in B, \alpha\in A}H^{\ri/\le}_{\beta', \alpha}(z,y+\i \Delta_{\beta'}0)\\[2mm]
&=-\sum_{\beta'\in B, \gamma\in C}H^{\ri/\le}_{\beta', \gamma}(z,y+\i \Delta_{\beta'}0)\\[2mm]
&=v^{\ri/\le}(z, y)\hbox{ in }D^{+}\cap D^{\ri/\le}\times Y_{2}.
\eea
\]
In the first line we use \eqref{papepresident.e3},  in the second  and fourth lines we use that $H_{\gamma,\gamma'}= - H_{\gamma', \gamma}$, in the third line the fact that $H_{\beta', \alpha}\in \mo_{\rm temp}((I^{\ri/\le}\times Y_{2})+ \i \Gamma_{\beta', \alpha}0)$ and the property of boundary values of holomorphic functions recalled in \ref{sec5.1.2},  and in last line \eqref{papepresident.e4} and \eqref{papepresident.e2}.

Summarizing we have:
\beq\label{e5.1}
\tilde{u}^{\ri/\le}(t, y)= \tilde{v}^{\ri/\le}(t+ \i 0, y)
\eeq
\beq\label{e5.0}
\begin{array}{l}
P_{z}\tilde{v}^{\ri/\le}(z, y)= w(z, y)\hbox{ in }D^{+}\times Y_{2}, \\[2mm]
\tilde{v}^{\ri/\le}(z, y)= v^{\ri/\le}(z, y) \hbox{ in }(D^{+}\cap D^{\ri/\le})\times Y_{2}.
\end{array}
\eeq

{\it Step 3.}

We  now set
\beq\label{e5.2}
\tilde{v}= \tilde{v}^{\ri}- \tilde{v}^{\le}\in \mo_{\rm temp}(D^{+}; \cD'(Y_{2})).
\eeq
From \eqref{e5.-10}, \eqref{e5.0} we obtain
\beq\label{e5.3}
\bea
P_{z}\tilde{v}&=0\hbox{ in }D^{+}\times Y_{2}, \\
\chi v(s, y)&= \tilde{v}(\i s, y)\hbox{ in }I^{+}\times Y_{2}.
\eea
\eeq
Let
\beq\label{e5.3b}
\tilde{u}(t, y)= \tilde{v}(t+ \i 0, y)\in \cD'(I\times Y_{2}).
\eeq
From \eqref{e5.3} we have $P_{t}\tilde{u}(t, y)=0$ in $I\times Y_{2}$ and $\WFA(\tilde{u})\subset \{\tau\geq 0\}$ over $I\times Y_{2}$.  By microlocal ellipticity (see \cite[Corr. 2.1.2]{SKK} or \cite[Thm. 5.1]{H4}) this implies that $\WFA(\tilde{u})\subset \cN^{+}$ over $I\times Y_{2}$.

{\it Step 4.}

We claim that
\begin{equation}
\label{e5.4}
\rho_{\Sigma} \tilde{u}= g\hbox{ on } Y_{2}.
\end{equation}
Note that from  \eqref{e5.4} and finite propagation speed one can conclude that $\tilde{u}= U_{\Sigma}g$ near $I\times Y_{3}$ for $Y_{3}\Subset Y_{2}$. Since $\WFA(\tilde{u})\subset \cN^{+}$ over $I\times Y_{2}$ we obtain the proposition, by the discussion in Step 1. 

Therefore it remains to check \eqref{e5.4}. We recall that the cutoff functions  $\tilde{\psi}\in \coinf(Y)$, $\chi\in \coinf(I\times Y)$ were introduced in Step 1 and that $\tilde{\psi} g= \gamma^{+} \chi v$. Since $P_{t}\tilde{u}=0$ in $I\times Y_{2}$ and $\{t=t_{0}\}$ are non-characteristic for $P_{t}$ for $t$ close to $t_{0}$, we know that $\tilde{u}\in \cinf(I; \cD'(Y_{2}))$. For $\varphi\in \cD(Y_{2})$ we set
\[
\begin{array}{l}
\tilde{u}_{\varphi}(t)= \langle \tilde{u}(t, \cdot), \varphi(\cdot)\rangle_{Y}\in\cinf(I),\\[2mm]
\tilde{v}_{\varphi}(z)= \langle \tilde{v}(z, \cdot), \varphi(\cdot)\rangle_{Y}\in \mo_{\rm temp}(D^{+}).
\end{array}
\]
By \eqref{e5.3b} we have
\[
\tilde{u}_{\varphi}(t)= \lim_{\epsilon\to 0^{+}}\tilde{v}_{\varphi}(t+ \i \epsilon), \hbox{ in }\cD'(I).
\]
Since $\tilde{u}_{\varphi}\in \cinf(I)$ we deduce then from  \cite[Thm. 3.6]{Kom} that 
\[
\tilde{u}_{\varphi}(t)= \lim_{\epsilon\to 0^{+}}\tilde{v}_{\varphi}(t+ \i \epsilon) \hbox{ in }\cinf(I).
\] 
In particular  by \eqref{e5.3} we have 
\[
\tilde{u}_{\varphi}(0)=  \lim_{\epsilon\to 0^{+}}\langle \chi v(\i \epsilon, \cdot), \varphi(\cdot)\rangle_{Y}=  \langle \gamma^{+}_{0} \chi v, \varphi\rangle_{Y},
\]
which by \eqref{e5.0ER} implies that
\[
\tilde{u}(0, y)= g_{0}(y)\hbox{ in }\cD'(Y_{2}),
\]
for $g= \col{g_{0}}{g_{1}}$. The same argument for $\p_{t}\tilde{u}$ shows that $\i^{-1}\p_{t}\tilde{u}(0, y)= g_{1}(y)$ in $\cD'(Y_{2})$. This completes the proof of \eqref{e5.4}. \qed

\begin{proposition}
The  state $\omega_{\Omega}$ constructed in Thm. \ref{turlututu} is an analytic Hadamard state.
 \label{propito}
\end{proposition}
\proof We recall that the spacetime covariances of $\omega_{\Omega}$ are given by:
\[
\Lambda^{\pm}= (\rho_{\Sigma}\circ G)^{*} \circ \lambda^{\pm}\circ (\rho_{\Sigma}\circ G),
\]
for  $\lambda^{\pm}= q\circ C_{\Omega}^{\pm}$.
Since  the solution of the Cauchy problem
\[
\begin{cases}
P\phi= 0,\\
\rho_{\Sigma}\phi=f
\end{cases}
\]
is given by:
\[
\phi= U_{\Sigma}f= G^{*}\circ \rho^{*}_{\Sigma}\circ qf,
\]
we obtain 
\[
\Lambda^{\pm}= U_{\Sigma}\circ C_{\Omega}^{\pm}\circ (\rho_{\Sigma}\circ G).
\]
Denoting by $x= (t, y)$ the variables in $M= \rr\times \Sigma$ we obtain that the distribution kernel 
  $\Lambda^{\pm}(x, x')$ solves the equation:
\[
\begin{cases}
P(x, \p_{x})\Lambda^{\pm}(x, x')=0 \hbox{ in  }M\times M,  \\
\rho_{\Sigma} \Lambda^{\pm}(\cdot, x')(\rx)= r^{\pm}(y, x'),
\end{cases}
\]
where $r^{\pm}(y, x')\in \cD'(\Sigma\times M)^{2}$ is the distribution kernel   of $(C_{\Omega}^{\pm}\circ \rho\circ G)$.

We can now repeat verbatim the arguments in the proof of Prop. \ref{prop5.1}, replacing $\Sigma$ by $\Sigma\times M$, the extra variable $x'$ playing simply the role of a parameter. We obtain  that
\[
\WFA(\Lambda^{\pm})'\subset \cN^{\pm}\times T^{*}M.
\]
Since $\Lambda^{\pm}$ is hermitian, this also implies that $\WFA(\Lambda^{\pm})'\subset T^{*}M\times \cN^{\pm}$ hence $\WFA (\Lambda^{\pm})'\subset \cN^{\pm}\times \cN^{\pm}$. \qed

\subsection*{Acknowledgments} The authors would like to thank Pierre Schapira for all the useful discussions. Support from the grants ANR-12-BS01-012-01 and ANR-16-CE40-0012-01 is gratefully acknowledged.

\appendix
\section{}\init\label{secapp1}
\subsection{An auxiliary lemma}\label{ssecap1}

\begin{lemma}\label{lem:app1} Let $v\in \cE'(\Omega)$ be equal to $\delta_{s}\otimes f$ or $\delta'_{s}\otimes f$ with $\pm s\leq 0$, $f\in C_{\rm c}^\infty(\Sigma)$. Then
\begin{equation}
\label{b5.11}
r^{\pm}K_{0}^{-1} v \in \overline{H^{1}_{0}}(\Omega^{\pm})\cap \overline{\cinf}(\Omega^{\pm}).
\end{equation}
\end{lemma}
\proof 
Let $\chi, \chi_{1}, \chi_{2}\in \coinf(\Omega)$ be cutoff functions with $\chi= 1$ near $\supp v$, $\chi_{1}= 1$ near $\supp \chi$ and $\chi_{2}= 1$ near $\supp \chi_{1}$.  We then have:
\beq\label{b5.10}
\bea
r^{\pm}K_{0}^{-1} v&= r^{\pm}\chi_{2}K_{0}^{-1}\chi  v+ r^{\pm}(1-\chi_{2})K_{0}^{-1}\chi  v\\
&= r^{\pm}\chi_{2}K_{0}^{-1}\chi  v+ r^{\pm}(1-\chi_{2})K_{0}^{-1}[\chi_{1}, K_{0}]K_{0}^{-1}\chi  v.
\eea
\eeq
By assumption, $\chi  v\in H^{-2}_{\rm c}(\Omega)$. By elliptic regularity $K_{0}^{-1}: H^{s}_{\rm c}(\Omega)\to H^{s+2}_{\rm loc}(\Omega)$, therefore $[\chi_{1}, K_{0}]K_{0}^{-1}\chi  v\in H^{-1}_{\rm c}(\Omega)$.  By the definition of $K_{0}^{-1}$ via quadratic forms, we know that $K_{0}^{-1}: H^{-1}(\Omega)\to H^{1}_{0}(\Omega)$, hence  
\[
r^{\pm}(1- \chi_{2})K_{0}^{-1}[\chi_{1}, K_{0}]K_{0}^{-1}\chi  v\in \overline{H^{1}_{0}}(\Omega^{\pm}).
\]
On the other hand, by elliptic regularity we know that $(1- \chi_{2})K_{0}^{-1}[\chi_{1}, K_{0}]$ is infinitely smoothing, hence
\[
r^{\pm}(1- \chi_{2})K_{0}^{-1}[\chi_{1}, K_{0}]K_{0}^{-1}\chi  v\in \overline{\cinf}(\Omega^{\pm}).
\]
Let us now consider the 
 first term in the second line of \eqref{b5.10}. By Lemma \ref{l4.1} we know  that $\chi_{2}K_{0}^{-1}\chi= \chi_{2}Q\chi+ R_{-\infty}$, where $Q\in \Psi^{-2}_{\rm c}(\Omega)$ and $R_{-\infty}$ has a smooth, compactly supported kernel in $\Omega\times \Omega$.  The term
 \[
r^{\pm}R_{-\infty} v
\]
 obviously belongs to $\overline{\coinf}(\Omega^{\pm})$.  Next, from \cite[Thm. 10.25]{Gr}   we know that $\chi_{2}Q\chi(\delta_s\otimes\cdot)$ and $\chi_{2}Q\chi(\delta'_s\otimes\cdot)$ are {\em Poisson operators}. In particular by \cite[Thm. 10.29]{Gr}, $\chi_{2}Q\chi(\delta_s\otimes\cdot)$ and $\chi_{2}Q\chi(\delta'_s\otimes\cdot)$ map $\coinf(\Sigma)^{2}$ into $\overline{\coinf}(\Omega^{\pm})$ continuously. Therefore,
\[
r^{\pm}\chi_{2}K_{0}^{-1}\chi  v\in \overline{\coinf}(\Omega^{\pm}).
\]
In conclusion we get \eqref{b5.11}. \qed


 \end{document}